\documentclass[galaxies,review,accept,oneauthor,pdftex]{Definitions/mdpi} 

\firstpage{1} 
\pubvolume{xx}
\issuenum{1}
\articlenumber{5}
\pubyear{2020}
\copyrightyear{2020}
\history{Received: date; Accepted: date; Published: date}
\updates{yes} 

\usepackage{amssymb}

\Title{
{Leptonic and Hadronic Radiative Processes in Supermassive-Black-Hole Jets}
}

\Author{{Matteo Cerruti} 
 \orcidA{}}

\AuthorNames{Matteo Cerruti}

\address[1]{Institut de Ci\`{e}ncies del Cosmos (ICCUB), Universitat de Barcelona (IEEC-UB), Mart\'{i} i Franqu\`{e}s 1, E08028~Barcelona, Spain; matteo.cerruti@icc.ub.edu
}

\abstract{Supermassive black holes lying in the center of galaxies can launch relativistic jets of plasma along their polar axis. The physics of black-hole jets is a very active research topic in astrophysics, owing to the fact that many questions remain open on the physical mechanisms of jet launching, of particle acceleration in the jet, and on the radiative processes. In this work I focus on the last item, and present a review of the current understanding of radiative emission processes in supermassive-black-hole jets.}


\begin{document}

\section{Introduction}
\unskip
\vspace{-10pt}
\subsection{Active Galactic~Nuclei}
\begin{figure}[H]
\centering
\includegraphics[width=16 cm]{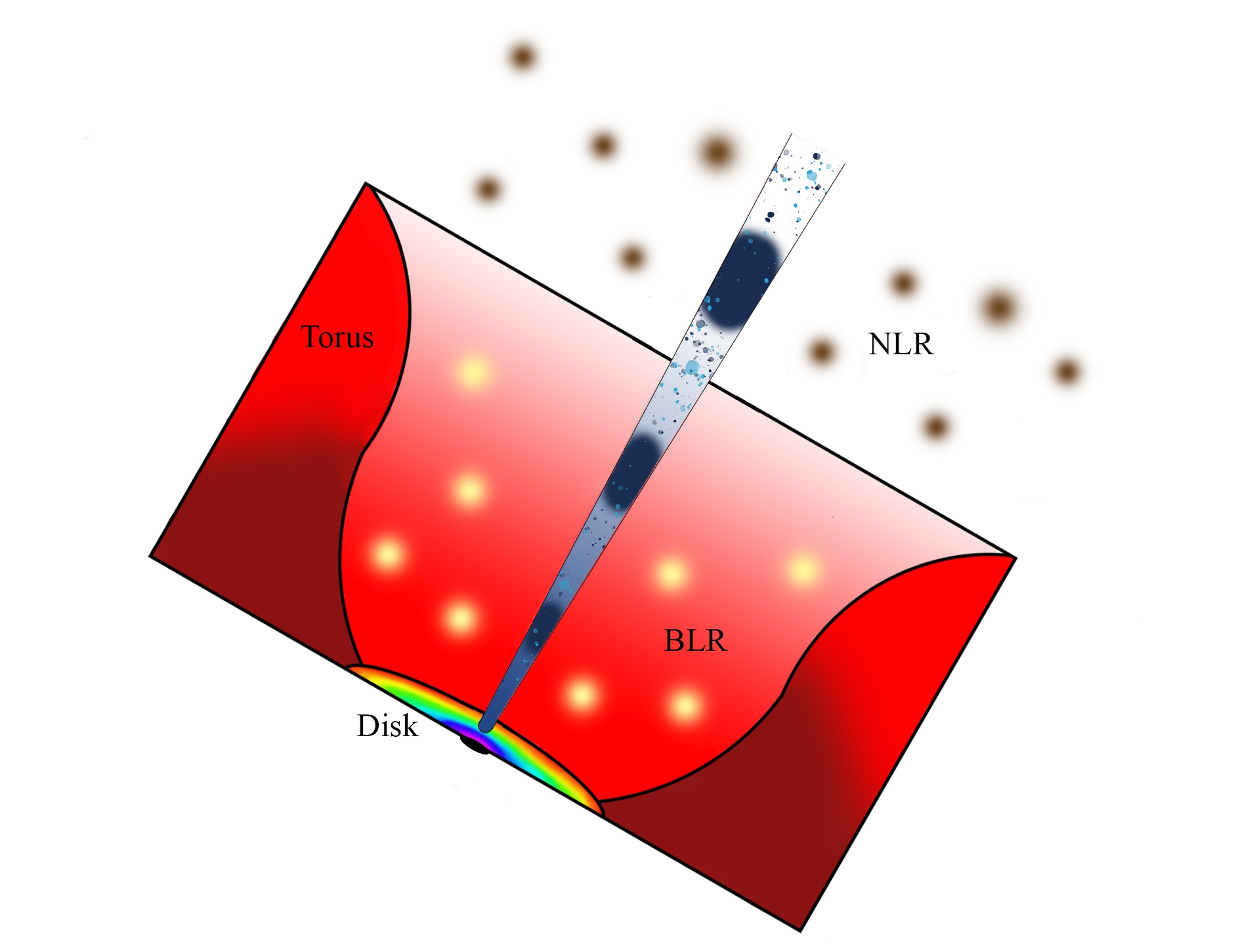}
\caption{Schematic representation (not in scale) of the central regions of a radio-loud~AGN. 
 \label{figone}
}
\end{figure}

The brightest, persistent objects in the Universe are active galactic nuclei (AGNs), compact regions in the center of galaxies that can outshine the host itself (composed of hundreds of billions of stars), and~can be detected on cosmological distances, up~to a redshift $z = 7.5$ \citep{Banados18}, which means a comoving distance of about 9 Gpc in the current $\Lambda$CDM cosmological model \citep{Planck19}. About 75 years of observations at all wavelengths have collectively shaped the so-called AGN unified model: the engine powering the system is a supermassive black hole
(SMBH) of $10^{8-9}\ M_\odot$ \citep{EHTpaper1}, which accretes matter in the form of an accretion disk \citep{Abramowicz2013} and which is surrounded at the parsec scale by an obscuring dusty torus \citep{Jaffe93}. 
The thermal emission from the disk ionizes clouds of matter orbiting the SMBH, which then re-emit emission lines in optical/UV (the broad and narrow line-regions, BLR/NLR, the~first one close to the black hole at the sub-parsec scale \citep{Greene05, Bentz09}, the~second one at much larger distances, hundreds of parsecs and more \citep{Bennert02}, beyond~the torus, see a schematic representation in Figure \ref{figone}).
The inclination of the system with respect to the observer is responsible for the variety of AGN types, due to the fact that when the system is seen face-on the observer has direct access to the SMBH, while when the system is seen edge-on the inner regions are obscured by the torus \citep{Netzer15}. Quasars (quasi-stellar objects) are now understood as luminous AGNs seen face-on. The~host galaxy of the first identified quasars (such as 3C 273 \citep{Schmidt63}) can now be resolved and studied with current instrumentation \citep{Bahcall97}.

A major dichotomy exists among AGNs: a minority of around 10$\%$ of AGNs are bright in the radio band \citep{Wilson95} and are hence dubbed \textit{radio-loud} AGNs, in~contrast with the more common \textit{radio-quiet} AGNs. The~radio non-thermal emission is associated with a pair of relativistic jets of plasma which are launched along the polar axis of the SMBH and perpendicularly to the accretion disk. These jets can travel up to the Mpc scale, exceeding the (visible) size of the AGN host galaxy. For~the nearest and brightest objects, the~emission from the relativistic jet has been resolved not only in the radio band~\citep{Kellermann98}, but~also in optical~\citep{Butcher80}, X-rays \citep{Marshall05}, and~more recently in $\gamma$-rays \citep{LATCenA, HESSCenA}. AGNs with extended radio emission from the jets are called \textit{radio-galaxies}. In~this case the observer sees the jet from the side. When~the relativistic jet points instead in the direction of the observer, relativistic effects boost the emission \citep{Blandford78, Blandford79}, making~these AGNs particularly luminous in the Universe. These radio-loud AGNs observed down-the-jet are called \textit{blazars} \citep{Urry95}, and~are characterized by high luminosity, rapid~variability, and~high degree of polarization \citep{Angel80}. Although~blazars are a minority among AGNs their high luminosity makes them ubiquitous in all extragalactic surveys at all wavelengths. It is remarkable that in $\gamma$-rays, where~the radio-quiet AGNs do not emit anymore, around $70 \%$ of all the known extragalactic sources are blazars \citep{4FGL}. 

\subsection{Blazars}
Blazars come in two flavors: BL Lacertae objects, whose optical/UV spectrum is characterized by a featureless non-thermal continuum \citep{Shaw13}, and~Flat-Spectrum-Radio-Quasars (FSRQs), which~show instead broad emission lines from the BLR typical of quasars \citep{Shaw12}. This dichotomy is also observed in the radio-galaxies parent population, which are divided into Fanaroff–Riley type I radio-galaxies, with~strong core emission and relatively fainter jets, and~Fanaroff–Riley type II radio-galaxies, with~much stronger jets that terminate in a shock with the inter-galactic medium \citep{Fanaroff74}.

The fact that the relativistic jet of plasma moves towards the observer with speed $v=\beta c$, Lorentz~factor $\Gamma  = (1-\beta^2)^{-0.5}$, and~viewing angle $\vartheta_{view}$ has three important consequences coming directly from special~relativity: 
\begin{itemize}
[leftmargin=*,labelsep=4.9mm] 
\item plasma moving in the jet will be seen in the observer's frame as showing a \textit{projected} superluminal speed $\beta_{project} > 1$. It can be shown that $\beta_{project} = (\beta \sin\vartheta_{view}) / (1 - \beta \cos\vartheta_{view}) $, which can be as high as $\beta \Gamma$. Indeed, the~detection of superluminal motion in blazar jets is one of the key observations supporting the AGN unified model \citep{Mojave13}.
\item the emission from the jet is boosted in the direction of movement, which translates into a flux density in the observer's frame\footnote{Here and in the following, primed quantities refer to the jet's frame, while non-primed quantities to the observer's frame.} which scales as $F_\nu = \delta^3F^\prime_{\nu^\prime}$, where the quantity $\delta~=~[\Gamma(1-~\beta \cos \vartheta_{view})]^{-1}$ is defined as the \textit{Doppler factor}. Typical values for the Doppler factor in blazars are of the order of ten, which means that the flux density in the observer's frame is typically a factor of a thousand higher than in the jet's frame. Similarly, it means that if we have a radio-galaxy and a blazar at the same redshift, and~with identical flux densities in the jet's frame, the~blazar will be a thousand times brighter than the radio-galaxy as observed from~Earth. 
\item the effect of time-compression translates into a flux variability in the observer's frame much faster than the one in the jet's frame: the variability timescale, conventionally defined as the flux-doubling timescale $\tau_{var}$, is equal to $(1+z)\ \tau_{var}^\prime / \delta$. For~a typical blazar Doppler factor of ten, a~flare with a timescale in the jet's frame of one hour, is compressed in the observer's frame into a minutes-long flare.
\end{itemize}

The spectral energy distribution (SED) \footnote{We define SED the energy density representation $\nu F_{\nu}$, where $F_{\nu}$ is the flux density in units of erg cm$^{-2}$ s$^{-1}$ Hz$^{-1}$. It is easy to show that while  $F_{\nu}$ is different from $F_{\lambda}$ (the flux density per unit wavelength) and $F_{E}$ (the flux density per unit energy), the~SED is constant and $\nu F_{\nu} = \lambda F_\lambda = E F_E$.} of blazars is characterized by a non-thermal continuum extending from radio to very-high-energy $\gamma$-rays, with~only subdominant thermal emission in optical/UV associated with the accretion disk (the \textit{big-blue-bump}) and emission lines from the BLR (only seen in FSRQs, and~not in BL Lacs). This non-thermal continuum is comprised of two distinct radiative components: the first one peaks in infrared-to-X-rays, the~second one peaks in the $\gamma$-ray band, ranging from MeV to TeV. A~remarkable discovery by $\gamma$-ray telescopes is that in FSRQs the majority of the energy is emitted in the $\gamma$-ray band \citep{Hartman92}, and~in BL Lacs the energy emitted in the $\gamma$-ray band is as much as the one emitted in the radio-to-X-ray band \citep{Punch92}. The~SED peak frequencies are not fixed, not within the blazar population, nor in a single blazar. FSRQs show typically a first SED peak in the infrared band, and~a second SED peak in the MeV band; BL Lacs, on~the other hand, show a huge variety of peak frequencies, from~infrared to X-rays, and~from MeV to TeV. The~peak frequency is used to further classify BL Lacs \citep{Padovani95, FermiSED} into low-frequency-peaked BL Lacs (LBLs, with~$\nu_{peak} < 10^{14}$ Hz, in~the infrared band, similarly to FSRQs) and high-frequency-peaked BL Lacs (HBLs, with~$\nu_{peak} > 10^{15}$ Hz, in~the UV/X-ray band). The~transition among the two sub-classes is smooth, and~there exist intermediate-frequency-peaked BL Lacs (IBLs, with~$10^{14}$ Hz $< \nu_{peak} < 10^{15}$ Hz, in~optical, \citep{Laurent99}).  It is not clear how the peak frequency distribution ends: observations by Cherenkov telescopes in the TeV band have identified a population of extremely high-frequency-peaked blazars (EHBLs) with  $\nu_{peak} > 10^{18}$ Hz, in~the hard-X-rays \citep{Costamante01, Bonnoli15}, and, correspondingly, a~high-energy SED peak extending into the TeV band \citep{HESS0229}. More recent studies seem to indicate however that the EHBL population is not homogeneous, with~all various combinations of soft/hard X-rays and soft/hard TeV spectra possible \citep{Foffano19, Biteau20, Costamante20}.
It seems that the peak frequency is anti-correlated with the blazar luminosity: the brightest blazars are the ones with the lowest $\nu_{peak}$, while the faintest ones are the ones with the highest $\nu_{peak}$. This anti-correlation defines the \textit{blazar sequence} \citep{Fossati98, Meyer11, Ghisellini17}. The~existence of this sequence is still an open question in the blazar research field, and~several outliers to the sequence have been identified \citep{Padovani03, Nieppola08, Padovani12, Giommi12}. If~the blazar sequence represents a true physical property of blazars, it will help understanding what physical processes shape the blazar SEDs \citep{Bott02, Ghisellini08, Finke13}.

The framework described up to here is pretty solid, and~has been tested against a variety of observations at all wavelengths.  Nonetheless, there is still many questions which remain open. Only limiting the discussion to radio-loud AGNs, it is still not completely understood \textit{{(i)}
} how relativistic jets are formed and powered, and~why only some AGNs can produce them; \textit{{(ii)}} why there are two intrinsically different jet types as seen in FRI/FRII and BL Lacs/FSRQs; \textit{{(iii)}} how particles are accelerated in jets; \textit{{(iv)}} how photons are produced in jets.
The last item is the subject of this review. I~will summarize the state-of-the-art answer to the following question: \textit{how do supermassive-black-hole jets~shine?} For a review of recent progress on observational results see \citep{Blandfordreview, Hovattareview}. For~a broader discussion of relativistic jets at all scales (AGNs and micro-quasars) see \citep{Romero17}.

The blazar non-thermal continuum emission implies the existence of a non-thermal population of particles that can radiate photons over the whole electromagnetic spectrum. Leptonic radiative processes are the ones associated with electrons/positrons in the jet; hadronic radiative processes are the ones associated with protons and nuclei. In~the following I will discuss how photons (and as it will be shown later, neutrinos) can be emitted from jets, assuming that there exists an efficient particle accelerator in the jet. For~particle acceleration in AGN jets, see \citep{Kirk00, Sironi15}

\section{Leptonic Radiative~Processes}
\unskip
\vspace{-10pt}
\subsection{Electron Synchrotron~Emission}
\label{elsynsubsec}

Synchrotron radiation is produced by charged particles moving in a magnetic field. For~sake of simplicity, let's start with the assumption that the emitting region is a sphere of radius $R$ in the jet, moving with bulk Lorentz factor $\Gamma$, and~filled with a tangled, homogeneous magnetic field $B^\prime$. If~a particle with electric charge $e$ and Lorentz factor $\gamma^\prime = E^\prime/mc^2$ is put in the region, it radiates a synchrotron power (in units of erg  s$^{-1}$  Hz$^{-1}$) per frequency $\nu^\prime$ equal to \footnote{For further details on this and other radiative mechanisms see \citep{BlumenthalGould, Rybicki79, Longair94, Dermer09}.}
\begin{linenomath}\begin{equation}
P^\prime_{syn, \nu^\prime} (\nu^\prime,  \gamma^\prime, \varphi^\prime) = \frac{\sqrt{3} e^3B^\prime \sin{\varphi^\prime}}{m c^2} \frac{\nu^\prime}{\nu^\prime_c} \int^\infty_\frac{\nu^\prime}{\nu^\prime_c} {dx\ K_{5/3}(x)}
\end{equation}\end{linenomath} 
where $\varphi^\prime$ is the pitch angle between the particle momentum and the magnetic field lines, the~frequency $\nu^\prime_c$ is defined as $\nu^\prime_c(\gamma^\prime, \varphi^\prime) = \frac{3eB}{4\pi mc^2} \gamma^{\prime 2} \sin{\varphi^\prime}$, and~$K_{5/3}(x)$ is the modified Bessell function of the second kind of order $5/3$. The~sign of the charge does not matter: the synchrotron emission from electrons and positrons is not distinguishable, and~in the following I will simply talk about~electrons. 

Let's now assume that in the emitting region there is not one single electron, but~an isotropic population of electrons with a power-law distribution of Lorentz factors $\gamma^\prime_e$, defined between $\gamma^\prime_{min}$ and~$\gamma^\prime_{Max}$:

\begin{linenomath}\begin{equation}
\label{eq2}
N^\prime_e(\gamma^\prime_e) = N^\prime_0 \gamma_e^{\prime -n}
\end{equation}\end{linenomath}
where $N^\prime_0$ is the normalization of the electron distribution in units of cm$^{-3}$, and~$n$ the electron spectral index. The~collective synchrotron emissivity (in units of erg  s$^{-1}$ cm$^{-3}$  Hz$^{-1}$ sr$^{-1}$) from this population of electrons is then computed by integrating the synchrotron power over $\gamma^\prime_e$ and $\varphi^\prime$ as 

\begin{linenomath}\begin{equation}
j^\prime_{syn,\nu^\prime} (\nu^\prime) = \frac{1}{8\pi}\int_{\gamma^\prime_{min}}^{\gamma^\prime_{Max}}{d\gamma^\prime_e\ N^\prime_e(\gamma^\prime_e)}\ \int_{0}^{\pi}{d\varphi^\prime\ \sin{\varphi^\prime}\ P^\prime_{syn, \nu^\prime} (\nu^\prime,  \gamma^\prime, \varphi^\prime) } 
\end{equation}\end{linenomath}

Given the emissivity, the~$\nu F_\nu$ flux in the observer's frame (in units of erg cm$^{-2}$ s$^{-1}$) is computed~as

\begin{linenomath}\begin{equation}
\nu F_\nu (\nu)= \frac{4\pi}{3} \frac{R^3}{d_L^2} \delta^4 \nu^\prime j^\prime_{syn,\nu^\prime} (\nu^\prime) 
\end{equation}\end{linenomath}
where $d_L$ is the luminosity-distance of the source. It can be shown by performing analytically the integrals that if~the electron distribution is a power-law distribution with index $n$ as in Equation~(\ref{eq2}), then the differential photon flux $dN/dE$ (which is a quantity commonly used in high-energy astrophysics, equal to $F_E (E) / E$) is also a power-law distribution with index $\Gamma_{dN/dE} = (n+1) / 2$, the~flux density $F_\nu$ has index $\alpha = (n-1) / 2$, and~the energy flux  $\nu F_\nu (\nu)$  has index $p = (n-3) / 2$. Given that $\nu = \delta \nu^\prime / (1+z)$, and~that $j^\prime_\nu \propto \nu^{\prime -\alpha}$, the~energy flux is proportional to $\delta^{3+\alpha}$.\\

At low energies, an~important physical process is synchrotron self-absorption: the photons that the electrons radiate can be re-absorbed by the electrons themselves. Following Einstein's approach, it~is possible to calculate the self-absorption coefficient (in units of cm$^{-1}$) as 

\begin{linenomath}\begin{equation}
\label{selfabscoef}
\mu^\prime_{syn} (\nu^\prime) =  - \frac{1}{16\pi m\nu^2}\int_{\gamma^\prime_{min}}^{\gamma^\prime_{Max}}{d\gamma^\prime\ \gamma^{\prime 2} \frac{d}{d\gamma} \left( \frac{N^\prime_e(\gamma^\prime_e)} {\gamma^{\prime 2}} \right) }\ \int_{0}^{\pi}{d\varphi^\prime\ \sin{\varphi^\prime}\ P^\prime_{syn, \nu^\prime} (\nu^\prime,  \gamma^\prime, \varphi^\prime) } 
\end{equation}\end{linenomath}

In the presence of this absorption term, the~intensity of the synchrotron radiation (in units of erg~s$^{-1}$~cm$^{-2}$  Hz$^{-1}$ sr$^{-1}$) from this spherical emitting region in the jet can be computed as \citep{Bloom96}
\begin{linenomath}\begin{equation}
I^\prime_{syn,\nu^\prime} (\nu^\prime) = \frac{j^\prime_{syn,\nu^\prime} (\nu^\prime)}{\mu^\prime_{syn} (\nu^\prime)} \left\lbrace 1- \frac{2}{\tau (\nu^\prime)^2}\left[ 1 - \mathrm{e}^{-\tau (\nu^\prime)}(\tau(\nu^\prime)+1) \right] \right\rbrace
\end{equation}\end{linenomath}
where $\tau(\nu^\prime) = 2R \mu^\prime_{syn} (\nu^\prime)$. The~effect of self-absorption on the radiated synchrotron spectrum is a change in the power-law index: at frequencies below $\nu^\prime_{self-abs}$ the radiated flux density is proportional to $\nu^{5/2}$. Said otherwise, at~frequencies below $\nu^\prime_{self-abs}$ the emitting region is optically thick to synchrotron radiation.  Self-absorption in the emitting region has one important consequence: a single emitting region in the relativistic jet \textit{cannot} reproduce blazar SEDs in the radio band. The~emission in the radio band must come from other regions which are necessarily optically thin to synchrotron radiation, as~will be discussed in Section~\ref{extendedjet}.
 
\subsubsection{Self-Consistent Electron~Distribution} 
\label{sectioneqdiffeq}

As electrons radiate synchrotron photons, they lose energy, or, using the jargon of astrophysicists, cool. When calculating the synchrotron spectrum from a stationary electron population in the jet, it is thus important to first compute the self-consistent electron energy distribution at equilibrium. The~differential equation that describes this process~is  

\begin{linenomath}\begin{equation}
\label{equationdiffeq}
\frac{\partial}{\partial t} N^\prime_e(\gamma^\prime_e, t) = \frac{\partial}{\partial \gamma^\prime_e} \left[ \gamma^\prime_e \frac{N^\prime_e(\gamma^\prime_e, t)} {\tau_{syn}(\gamma^\prime_e)} \right] - \frac{N^\prime_e(\gamma^\prime_e, t)}{\tau_{ad}}  + Q^\prime_e(\gamma^\prime_e) 
\end{equation}\end{linenomath}
where $Q^\prime_e(\gamma^\prime_e)$ is the function describing continuous (time-independent) injection of electrons, $\tau_{syn}(\gamma^\prime)$ is the synchrotron cooling timescale, and~$\tau_{ad}$ is the adiabatic timescale. The~latter is the timescale associated with the expansion of the emitting region as it travels along the jet, and~it is energy-independent. It is usually assumed to be $\tau_{ad} \simeq R/c$.  From~a mathematical point of view, the~term in $\tau_{ad}$ can also be interpreted as an energy-independent particle escape from the emitting region. The~synchrotron cooling timescale is defined as $\tau_{syn}(\gamma_e^\prime) = \left( \frac{1}{\gamma_e^\prime} \frac{\partial \gamma_e^\prime}{\partial t} \right)^{-1}$ and it is equal~to

\begin{linenomath}\begin{equation}
\tau_{syn}(\gamma^\prime_e) = \frac{3m_ec}{4\sigma_T u^\prime_B} \frac{1}{\gamma^\prime_e}
\end{equation}\end{linenomath}
where $u^\prime_B = B^{\prime 2} / 8\pi$ is the magnetic energy density. The~particle distribution at equilibrium can be computed putting Equation~(\ref{equationdiffeq}) equal to zero. The~integral solution is \citep{Inoue96}

\begin{linenomath}\begin{equation}
N^\prime_e(\gamma^\prime_e) = e^{-\gamma^\prime_{e, break} / \gamma^\prime_e} \frac{\gamma^\prime_{e, break}\ \tau_{ad}}{\gamma^{\prime 2}_e} \int_{\gamma^\prime_e}^\infty d\zeta Q^\prime_e(\zeta) e^{\gamma^\prime_{e, break} / \zeta}
\end{equation}\end{linenomath}
where $\gamma^\prime_{e, break} =  \frac{3mc^2}{4\sigma_T u^\prime_B R}$ is the value of  $\gamma^\prime_e$ that satisfies $\tau_{syn}(\gamma^\prime_e) =\tau_{ad}$. It can be shown that if the injection term is $Q^\prime_e(\gamma^\prime_e) = Q^\prime_0 \gamma^{\prime -q}_e$,   the~effect of synchrotron cooling on the particle distribution is a change in the spectral index, which softens and becomes equal to $q+1$ for particle Lorentz factors above $\gamma^\prime_{e, break}$. From~the point of view of the energy flux distribution, it also softens above $\nu^\prime_{e, break}$, and~becomes equal to $p+1/2$.  This is true only if $q \geq 2$. If~$q < 2$, the~particle distribution shows a pile-up at $\gamma^\prime_{e, break}$ \citep{Kardashev62, Sauge}. Another specific case happens when the injection function is also characterized by a minimum Lorentz factor  $\gamma^\prime_{min, inj}$. If~$\gamma^\prime_{e, break} > \gamma^\prime_{min, inj}$ (slow-cooling regime), then~the solution is the same as before. If~however $\gamma^\prime_{e, break} < \gamma^\prime_{min, inj}$ the steady-state solution is completely cooled (fast-cooling regime), and~characterized by an index $q+1$ above $\gamma^\prime_{min, inj}$, and~an index of $2$ between $\gamma^\prime_{e, break}$ and $\gamma^\prime_{min, inj}$ \citep{Sari01}. 

We are now in the position to quantitatively compute a self-consistent synchrotron spectrum from an electron distribution in a spherical emitting region in the relativistic jet. The~SED can be parametrized by a broken-power-law distribution, proportional to $\nu^{7/2}$ below the self-absorption frequency $\nu_{self-abs}$, to~$\nu^{-p}$ between $\nu_{self-abs}$ and $\nu_{break}$, and~to $\nu^{-(p+1/2)}$ above $\nu_{break}$.

It would be surprising if such a simple scenario would be able to reproduce blazar SEDs, and~indeed, it does not. Electron synchrotron radiation from a single emitting region in the jet can reproduce the first blazar SED component, at~one price: while the emission is consistent with a broken-power-law distribution, the~electron indexes $n_{1,2}$ below and above $\gamma^\prime_{e, break}$ are not consistent with a simple break of 1 as calculated above. What we usually observe are much stronger breaks, with~$n_2 - n_1 > 1$, with~$n_1 = 1.5-2.5$ and $n_2 = 3.5-4.5$. For~this reason, the~most common approach is to consider $n_{1,2}$ and $\gamma_{e,break}^\prime$ as free independent model parameters \citep{Tavecchio10, Zhang12}, and~discuss a posteriori their consistency with a synchrotron cooling break. In~a minority of cases, observations are consistent with a broken power-law with a spectral break of one \citep{0152HESS, 0033MAGIC, 1727VERITAS, 2005HESS}. Alternatively, it is possible to assume that the electron distribution at equilibrium is in the fast-cooling regime. In~this case, if~$n_1$ is consistent with $2.0$, it is possible to fit the SED assuming an injection index $q \simeq 2.5-3.5$ \citep{1215VERITAS, Balokovic16}. This approach has however two drawbacks: the first one is that it needs an extra free parameter $\eta$ that is the scaling factor of the adiabatic timescale that has to be increased to be in the fast-cooling regime $\tau_{ad} = \eta R/c$; the second one is that an injection index $q\simeq 3.0$ is significantly softer than standard acceleration mechanisms. The~reason we observe stronger breaks may be that one of the assumptions we made are not correct: there could be multiple acceleration/emitting regions~\citep{Marscher14}, non-linear cooling terms (relevant when adding the inverse-Compton process, see Section~\ref{iC}), a~non-homogeneous magnetic field, incomplete cooling (i.e., the SED is from an electron population that is not at equilibrium)~\citep{Chiang02, Graff08}.
Alternatively, the~injected particle distribution may not follow a simple power-law function, and~could show intrinsic curvature in the form of a log-parabolic function~\citep{Massaro04, Tramacere09}, or~a power-law one with exponential cut-off at $\gamma^\prime_{Max}$. The~observational measurements of $n_1$ and $n_2$ may thus be the asymptotic power-law fits over a narrow energy band.

\subsubsection{Low-Energy SED~Modeling}
\label{extendedjet}

As discussed above, a~single emitting region in the jet radiating synchrotron photons cannot explain the observed spectrum from radio up to optical-X-rays, due to self-absorption in the radio band. Blazar SED modeling is thus, inevitably, at~least two-zones: a small, compact emitting region is responsible for the high-energy part of the first SED component (optical frequencies and above), the~one that shows rapid variability; a larger, less dense emitting region is where the radio photons are produced. It is natural to consider that this radio emission comes from the extended radio-jet. The~exact geometry is not clear: the radio emission can be produced downstream with respect to the optical-X-ray one, or~be associated with a second flow that encompasses the inner jet \citep{Sol89}, or~the whole emission {could} be produced not in a small overdensity in the jet, but~in a conical jet, with~varying magnetic field \citep{Ghisellini85, Potter12}. For~most blazars, a~two-zone model is enough to fit the first SED component~\citep{1943VERITAS}, although~the transition between the two components remains unclear, with~potential contribution of the extended region up to the optical/UV band \citep{2155HESS, MAGIC1722}.

A fit of the model to the observations can be used to put constraints on the physics of the emitting region, or~of the acceleration mechanisms. The~number of free parameters of the synchrotron model is nine: three for the emitting region ($\delta$, $B$, and~$R$), and~six for the particle distribution ($N_0$,~$\gamma^\prime_{min, break, Max}$,~and~$n_{1,2}$). If~we force a self-consistent electron distribution,  $\gamma^\prime_{break}$, and~$n_{2}$ can be expressed as a function of other parameters, and~the number of free parameters becomes seven; if~we further assume that $\gamma^\prime_{min, Max}$ are respectively low-enough ($\simeq$1) and high enough ($\geq$10$^6$) that they cannot be constrained by observations, we are left with five free parameters. Still, the~true  number of observables is much lower: the synchrotron spectrum is fully characterized by its peak frequency and luminosity, and~by its index, i.e.,~we have only three observables. This is true both for the small plasmoid in the jet, and~for the large component emitting in the radio band. We could constrain the model further if we could see the self-absorption frequency, but~the one for the plasmoid is hidden below the extended-jet emission. Synchrotron modeling of the low-energy SED of blazars is, by~itself, \textit{{degenerate}
}, which means that although it is possible to find a set of model parameters that can reproduce the SED, we cannot constrain the parameter~values. 

\subsubsection{Time-Dependent~Modeling}

Until now we focused on the synchrotron emission by a stationary distribution of electrons in the jet, comparing theoretical expectations with blazar SEDs, that can be seen as snapshots of the blazar emission at a given instant. One of the key observing properties of blazars is however variability, and~key pieces of physical information can be extracted by studying blazar flares. It is possible to simulate blazar light-curves by solving Equation~(\ref{equationdiffeq}) as a function of time, following the evolution of the electron distribution with time, and~the associated synchrotron~radiation.  

As a first approximation, a~synchrotron blazar flare can be modeled with an instantaneous injection of particles $Q^\prime_e(t^\prime) = Q_0\ \delta(t^\prime-t^\prime_0)$. In~this case the time-evolution is fully determined by the cooling timescale $\tau^\prime_{syn}$ for Lorentz factors higher than $\gamma^\prime_{break}$. The~measure of $\tau_{var}$ can thus provide a direct measurement of $\tau^\prime_{syn} = \tau^\prime_{syn} (B^\prime, \gamma^\prime_e)$ and thus help constrain the parameter~space.

Increasing complexity, all plausible acceleration processes are energy-dependent. In~a more realistic scenario we can thus define an acceleration timescale $\tau^\prime_{acc} = \tau^\prime_{acc} (\gamma^\prime_e)$. The~flare is thus expected to be characterized by an asymmetric profile, due to the different values of $\tau^\prime_{acc}$ and $\tau^\prime_{syn}$. The~energy dependency of both acceleration and cooling implies that the spectrum changes during the flare. In~a plot $\Gamma_{dN/dE}$ vs $F_{\nu}$ this {translates} into a hysteresis cycle. Equivalently, it will appear as a time-lag among light-curves computed at different energies \citep{Kardashev62, Mastichiadis97, Kirk98, Bottcher02}. Both hysteresis and time-lags have been observed in X-ray flare of blazars \citep{Takahashi94, Kataoka00, Zhang02, Sokolov04}, and~can be used to simultaneously constrain the magnetic field strength and the Doppler factor of the emitting region (with a degeneracy between the two), resulting in $B \simeq 0.1$ for $\delta \simeq 10$.  When modeling blazar flares, there is another timescale that needs to be taken into account that is the light-crossing time $R/c$: the variability will thus be diluted as observed from Earth if $R/c > \tau^\prime_{syn}$ \citep{Chiaberge99}.

\subsection{Inverse-Compton~Emission}
\label{iC}

\textls[-15]{Compton scattering is the interaction between a charged particle and a photon. Following~Section~\ref{elsynsubsec}} on synchrotron radiation, let us start again with an electron. In~the direct Compton scattering a photon scatters over an electron at rest in an atom, and~loses energy in the interaction. The~photon wavelength after scattering changes as 
\begin{linenomath}\begin{equation}
\Delta \lambda = \frac{h}{m_ec} (1 - \cos \zeta)
\end{equation}\end{linenomath}
with $\zeta$ the scattering angle, and~with $\frac{h}{m_ec}=\lambda_C$ the Compton~wavelength.

 In the inverse-Compton scattering a high-energy electron interacts with a low-energy photon, and~transfers part of its energy to it. The~inverse-Compton scattering is a very efficient process to produce X-ray and $\gamma$-ray photons in astrophysical sources starting from low-energy photon fields. Let us mark with a star the reference frame at rest with the electron, and let’s mark with the $1,2$~subscripts the values before and after scattering, respectively. The~photon a-dimensional energies are $\varepsilon^\star_{1,2}~=~h\nu^\star_{1,2} / m_ec^2$.  The~electron Lorentz factors are $\gamma^\star_{e;1,2} = E^\star_{e; 1,2} / m_ec^2$, with~$\gamma^\star_{e;1} = 1$. It can be shown with simple kinematic and energy/momentum conservation equations that in the reference frame at rest with the electron
\begin{linenomath}\begin{equation}
\label{ICeq1}
\varepsilon^\star_2 = \frac{\varepsilon^\star_1}{1+\varepsilon^\star_1(1-\cos\xi^\star)}
\end{equation}\end{linenomath}  
where $\xi^\star$ is the angle between {the incoming and scattered directions of the photon}. When $\varepsilon^\star_{1} \ll 1$, Equation~(\ref{ICeq1}) simplifies as $ \varepsilon^\star_2 \simeq \varepsilon^\star_1 $, which is the photon energy in the electron rest frame is unchanged. The~condition $\varepsilon^\star_{1} \ll 1$ represents the Thomson regime, where the photon energy is much lower than the electron rest mass. The~equations describing the change of reference frame from and to the observer (unstarred) are:
\begin{linenomath}\begin{equation}
\varepsilon^\star_1 = \varepsilon_1 \gamma_e (1 - \beta_e \cos \vartheta_1) 
\end{equation}\end{linenomath}
\begin{linenomath}\begin{equation}
\label{ICeq3}
\varepsilon_2 = \varepsilon^\star_2 \gamma_e (1 + \beta_e \cos \vartheta^\star_2) 
\end{equation}\end{linenomath}
where $\vartheta_1$ is defined as the angle between {the electron and the incoming photon, in~the observer's frame,} and $\vartheta^\star_2$ {the one of the scattered photon, in~the electron's rest frame}. In~the Thomson regime, we thus have
\begin{linenomath}\begin{equation}
\varepsilon_2 \simeq \varepsilon^\star_1 \gamma_e (1 + \beta_e \cos \vartheta^\star_2) \simeq \varepsilon_1 \gamma^2_e (1 - \beta_e \cos \vartheta_1) (1 + \beta_e \cos \vartheta^\star_2)
\end{equation}\end{linenomath}

In the case of a relativistic electron with $\gamma_e  \gg 1$ and $\beta_e \simeq 1$, the~photon energy in the observer's frame after scattering can be as high as $\varepsilon_2 \simeq 4 \varepsilon_1 \gamma^2_e$. This occurs in the so-called head-on approximation, which is for $\cos \vartheta_1 = -1$ and $\cos \vartheta^\star_2 = 1$.\\

The expression of the inverse-Compton cross-section is equal to $\sigma_T$ in the Thomson regime, while~for the general case it has been calculated by \citet{Klein29}, and~is known as the Klein–Nishina formula. The~differential cross-section, in~the reference frame of the electron, is
\begin{linenomath}\begin{equation}
\label{KN}
\frac{d\sigma_C}{d\varepsilon_2^\star d\Omega^\star} = \frac{3\sigma_T}{16\pi}
\left( \frac{\varepsilon_2^\star}{\varepsilon_1^\star}  \right)^2 \left(\frac{\varepsilon_2^\star}{\varepsilon_1^\star} + \frac{\varepsilon_1^\star}{\varepsilon_2^\star} - \sin^2 \xi^\star \right)
\end{equation}\end{linenomath}
where $d\Omega^\star = d\varphi^\star d\cos \xi^\star$. The~integral cross-section is
\begin{linenomath}\begin{equation}
\sigma_C(\varepsilon_1^\star) = \frac{3\sigma_T}{8 \varepsilon_1^{\star 2}} \left( 4 + \frac{2 \varepsilon_1^{\star 2}(1+ \varepsilon_1^{\star})}{(1+2 \varepsilon_1^{\star})^2}  + \frac{ \varepsilon_1^{\star 2} - 2  \varepsilon_1^{\star} - 2}{ \varepsilon_1^{\star}} \ln(1+2 \varepsilon_1^{\star})  \right) 
\end{equation}\end{linenomath}

The cross-section drops steadily above $\varepsilon^\star_1 \gg 1$, and~asymptotically behaves as $\sigma_C(\varepsilon_1^\star)~\simeq~\frac{3}{8} \frac{\sigma_T}{\varepsilon^\star_1 } \left( \log(2\varepsilon^\star_1 ) + 1/2\right)$. The~regime $\varepsilon^\star_1  \gg 1$ is known as the Klein–Nishina regime.  Solving~Equations~(\ref{ICeq1}) and (\ref{ICeq3}) for the Klein–Nishina case, it is easy to show that in the best case~$\varepsilon_2 \simeq \gamma_e$. 

Let's discuss now the inverse-Compton emission from a distribution of electrons $N^\prime_e(\gamma^\prime_e)$ scattering a distribution of photons $n^\prime_{ph}(\varepsilon^\prime)$. The~computation of the inverse-Compton emissivity is analytically complex, requiring transformation of the Klein–Nishina cross-section from the electron rest frame to the jet frame. It has been investigated under several approximations (e.g., Thomson and Klein–Nishina approximation, head-on collisions, monochromatic photons). It is best solved numerically performing integrals over the electron and photon distributions times the cross-section (Equation (\ref{KN})). For~an isotropic distribution of electrons and photons, the~inverse-Compton emissivity can be expressed as:
\begin{linenomath}\begin{equation}
j^\prime_C(\nu^\prime) = \frac{h^2 \nu^\prime}{4\pi m_ec^2} \int d\varepsilon^\prime n^\prime_{ph}(\varepsilon^\prime) \int d\gamma^\prime N^\prime_e(\gamma^\prime_e)\ C(\varepsilon^\prime, \gamma^\prime_e, \nu^\prime)
\end{equation}\end{linenomath}
where $C(\varepsilon^\prime, \gamma^\prime_e, \nu^\prime)$ is the Jones' Compton kernel \citep{Jones68}.

As for the synchrotron radiation, the~fact that part of the electron energy is transferred to the photons imply that electrons are cooling in the process. It is thus important to include the inverse-Compton cooling losses when computing the self-consistent electron distribution at equilibrium (see Section~\ref{sectioneqdiffeq}). Let's start first with the Thomson regime: in this case the electron losses due to inverse-Compton scattering (for an isotropic photon field) take a form which is very similar to the synchrotron one, and~the associated timescale is
\begin{linenomath}\begin{equation}
\tau_{C, Th}(\gamma^\prime_e) = \frac{3mc}{4\sigma_T u^\prime_{ph}} \frac{1}{\gamma^\prime_e}
\end{equation}\end{linenomath}
where $u^\prime_{ph}$ is the energy density of the scattered photon field. When computing the electron distribution at equilibrium, it is just possible to simply use a single cooling timescale $\tau_{cooling}(\gamma^\prime_e) = \frac{3mc}{4\sigma_T (u^\prime_B + u^\prime_{ph})} \frac{1}{\gamma^\prime_e}$, and~the effect on the electron distribution is again a broken-power-law with indexes $n$ and $n+1$ below and above $\gamma^\prime_{break}$ respectively (in the {slow} cooling regime).

In the Klein–Nishina regime, the~inverse-Compton losses are also suppressed: if the inverse-Compton losses dominate over the synchrotron ones for high Lorentz factors (as could occur in the presence of bright external fields), the~result on the steady-state electron distribution  is a hardening at the highest energies \citep{Moderski05}.

\subsubsection{Synchrotron-Self-Compton}

Going back to the spheroidal plasmoid travelling with Lorentz factor $\Gamma$ in the SMBH relativistic jet, filled with both an homogeneous magnetic field $B$ and an electron population $N^\prime_e(\gamma_e^\prime)$, it is natural that the synchrotron photons will suffer inverse-Compton scattering over the same electrons population that produced them. This kind of emission is called \textit{synchrotron-self-Compton} (SSC), and~it is a very efficient process to transfer energy from particles to photons \citep{Gould79}.

In the Thomson regime, it can be shown analytically that if the electron distribution 
$N^\prime_e(\gamma_e^\prime)$ is a power-law with index $n$, not only the synchrotron energy flux has index $p = (n-3)/2$, but~the SSC energy flux has also index $p = (n-3)/2$. The~SSC emission is thus mirroring the synchrotron radiation at higher energies. It is thus natural to investigate SSC emission as the radiative process responsible for the high-energy SED component in blazars. Now let us investigate two key aspects: how~\textit{high} are these SSC energies, and~what is the relative luminosity $L_{SSC} / L_{syn}$ \citep{Tavecchio98}:
\begin{itemize}
[leftmargin=*,labelsep=4.9mm] 
\item Let's assume that the synchrotron peak frequency $\nu_{syn,peak}$ is directly related to $\gamma^\prime_{break}$, i.e., let’s assume that $n_1 < 3$ and $n_2 > 3$. In~this case
\begin{linenomath}\begin{equation}
\label{synpeak}
\nu_{syn,peak} = \frac{e}{2\pi m_ec} B^\prime\gamma^{\prime 2}_{break} \frac{\delta}{1+z}\ \textrm{Hz}
\end{equation}\end{linenomath}

The SSC peak frequency in the Thomson regime can be 
easily computed assuming that the electrons at $\gamma^\prime_{break}$ are the ones primarily scattering the photons at $\nu_{syn,peak}$, and~thus $\nu_{SSC,peak}~=~4/3\ \nu_{syn,peak} \gamma^{\prime 2}_{break}$. Assuming typical values for HBLs, i.e.,~a $\nu_{syn,peak} \simeq 0.1$ keV, in~the soft X-rays, and~$\gamma^\prime_{break} \simeq 10^{4}$ (see next Section),  $\nu_{SSC,peak}$ can thus reach 10 GeV. The~ratio $\nu_{SSC,peak} / \nu_{syn,peak}$ can be used to put a strong constraint on the model parameters, given~that
\begin{linenomath}\begin{equation}
\gamma^{\prime}_{break} = \left( \frac{3}{4}  \frac{\nu_{SSC,peak}} {\nu_{syn,peak}} \right)^{1/2}
\end{equation}\end{linenomath}

\item the total luminosities of the synchrotron and SSC components $L_{syn, SSC}$ can be expressed, in~a first approximation, as~$L_{syn, SSC} \simeq \nu^\prime_{syn,SSC;peak}\ L_{syn, SSC} (\nu^\prime_{syn,SSC;peak})$, which is their peak SED luminosities. The~ratio of the two SED peak fluxes depends only on the magnetic energy density and the synchrotron photons energy density $u^\prime_{ph, syn}$:
\begin{linenomath}\begin{equation}
\frac{ \nu_{SSC,peak}\ F_{SSC} (\nu_{SSC,peak})}{ \nu_{syn,peak}\ F_{syn} (\nu_{syn,peak})} =
\frac{ \nu^\prime_{SSC,peak}\ L_{SSC} (\nu^\prime_{SSC,peak})}{ \nu^\prime_{syn,peak}\ L_{syn} (\nu^\prime_{syn,peak})} = \frac{u^\prime_{ph, syn}}{u^\prime_B}
\end{equation}\end{linenomath}   
\end{itemize}

 In the Klein–Nishina regime on the other hand, the~SSC emission is suppressed, producing a major correction to the shape of the SSC component at higher energies. As~we increase the peak of the synchrotron emission, the~SSC peak frequency increases quadratically until the scattering enters into the Klein–Nishina regime, and~higher-energy photons do not contribute any more to the emission.
Even for the Klein–Nishina regime, simple relations between the frequencies and luminosities of the synchrotron and SSC components can be computed  \citep{Tavecchio98}. In is important to underline that the computation of best-fit solutions for the SSC model relies on a good estimate of the peak frequencies and fluxes. Given the wide variety of peak frequencies in the blazar population, these may fall in relatively unexplored energy bands.

When discussing the modeling of the low-energy SED component of blazars as synchrotron radiation (see Section~\ref{extendedjet}), we showed that the modeling of the synchrotron component alone is degenerate. This is not the case when we add the information from the SSC radiative component. We recall that the number of free parameters for a spherical plasmoid in the jet, filled with an electron population parametrized as broken-power-law function, is nine. Assuming that $\gamma^\prime_{min, Max}$ are respectively low and high enough that they cannot be constrained by the data, we are left with seven model parameters. With~the information from both SED components, the~number of observables is six: the frequencies and luminosities of the synchrotron and SSC peaks, and~the two indexes of the synchrotron distribution before and after the peak. We are left with just one missing constraint, which can be the self-consistency of the electron distribution, or~can be the observed fastest variability timescale $\tau_{var}$. It can be related to the size of the emitting region $R$ and the Doppler factor $\delta$ via a causality argument
\begin{linenomath}\begin{equation}
\label{causalityequation}
R \leq \frac{\tau_{var}}{c} \frac{\delta}{(1+z)}
\end{equation}\end{linenomath}

\begin{figure}[H]
\centering
\includegraphics[width=7.75cm]{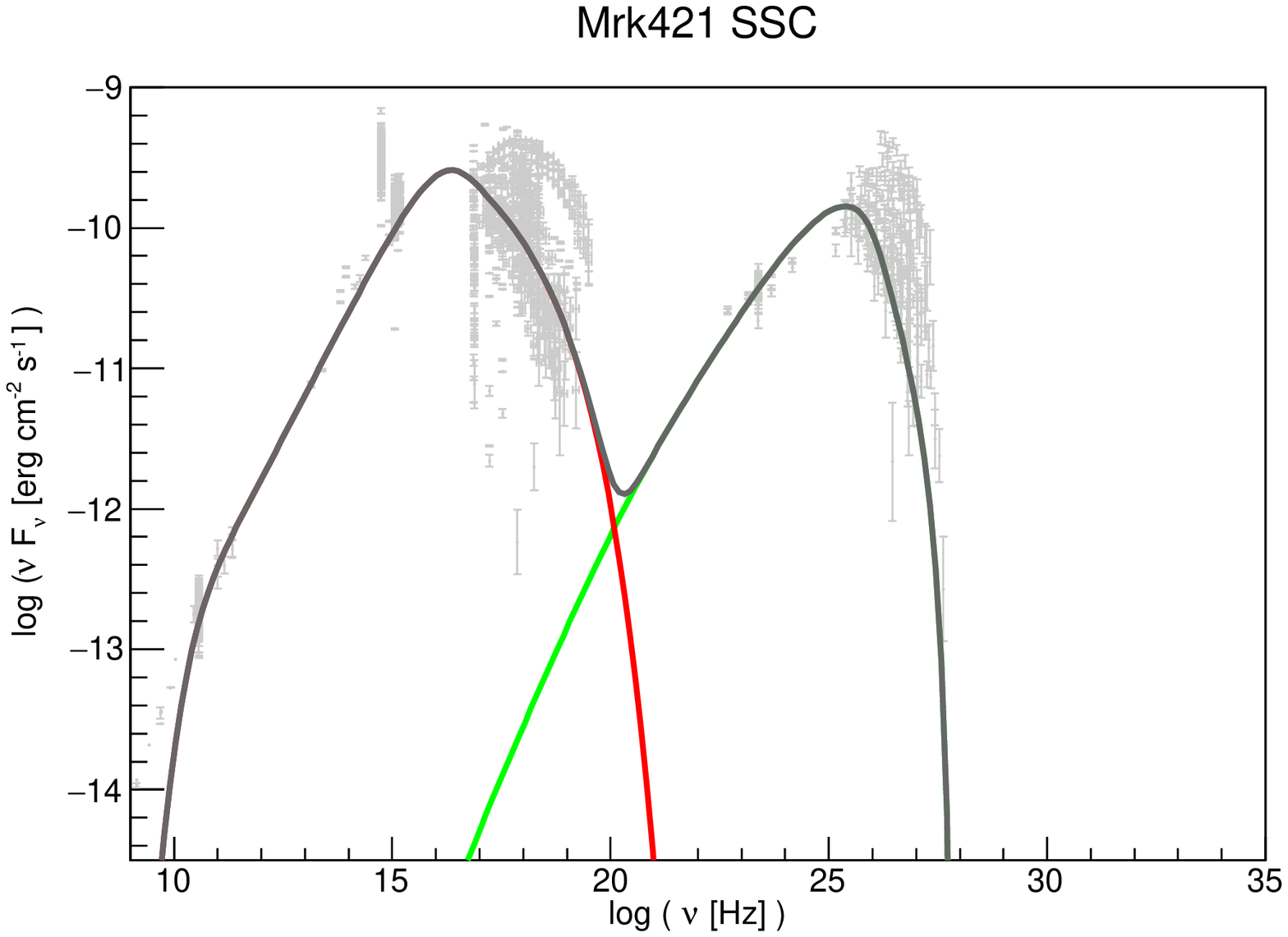}
\includegraphics[width=7.75cm]{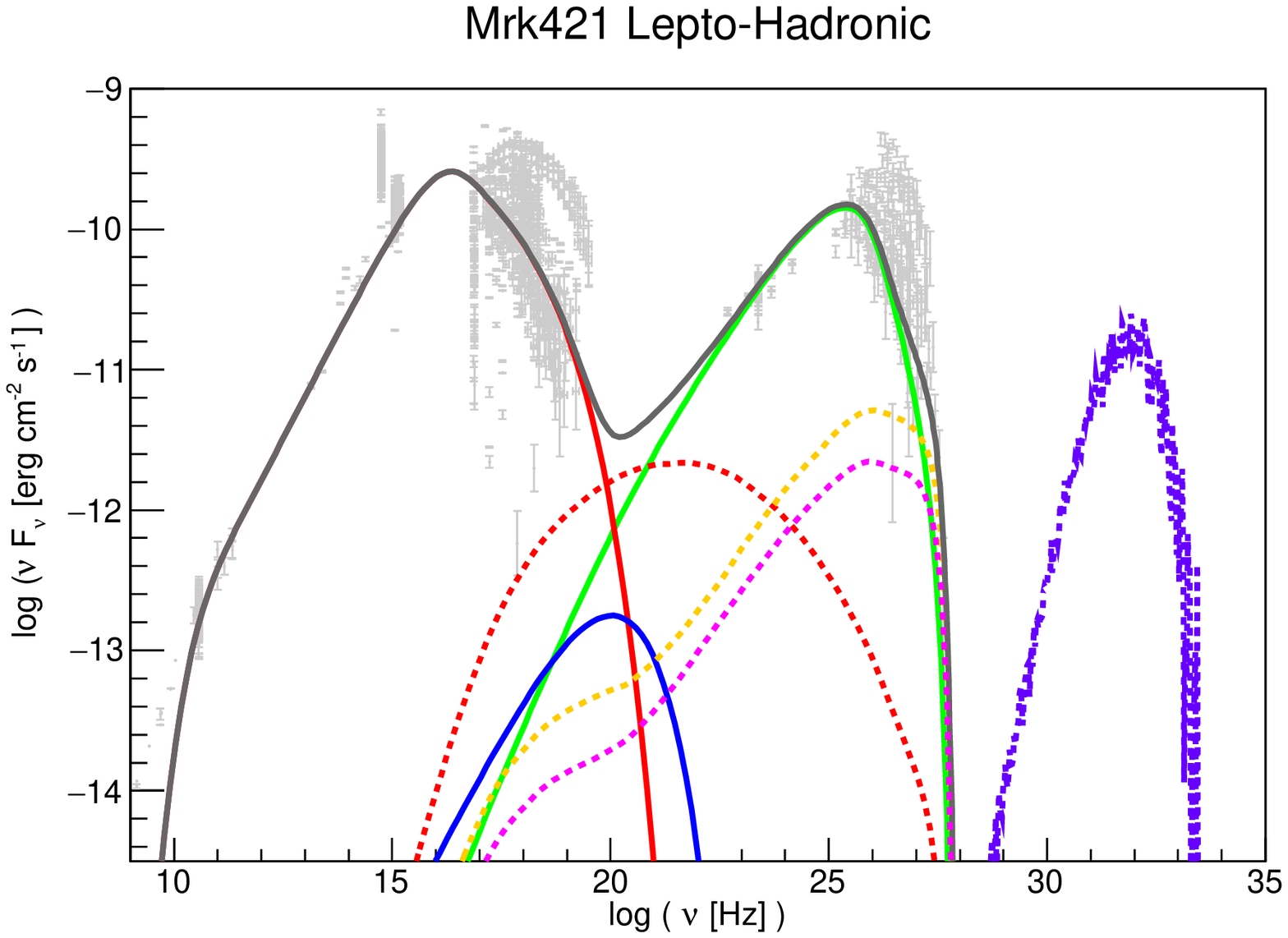}
\includegraphics[width=7.75cm]{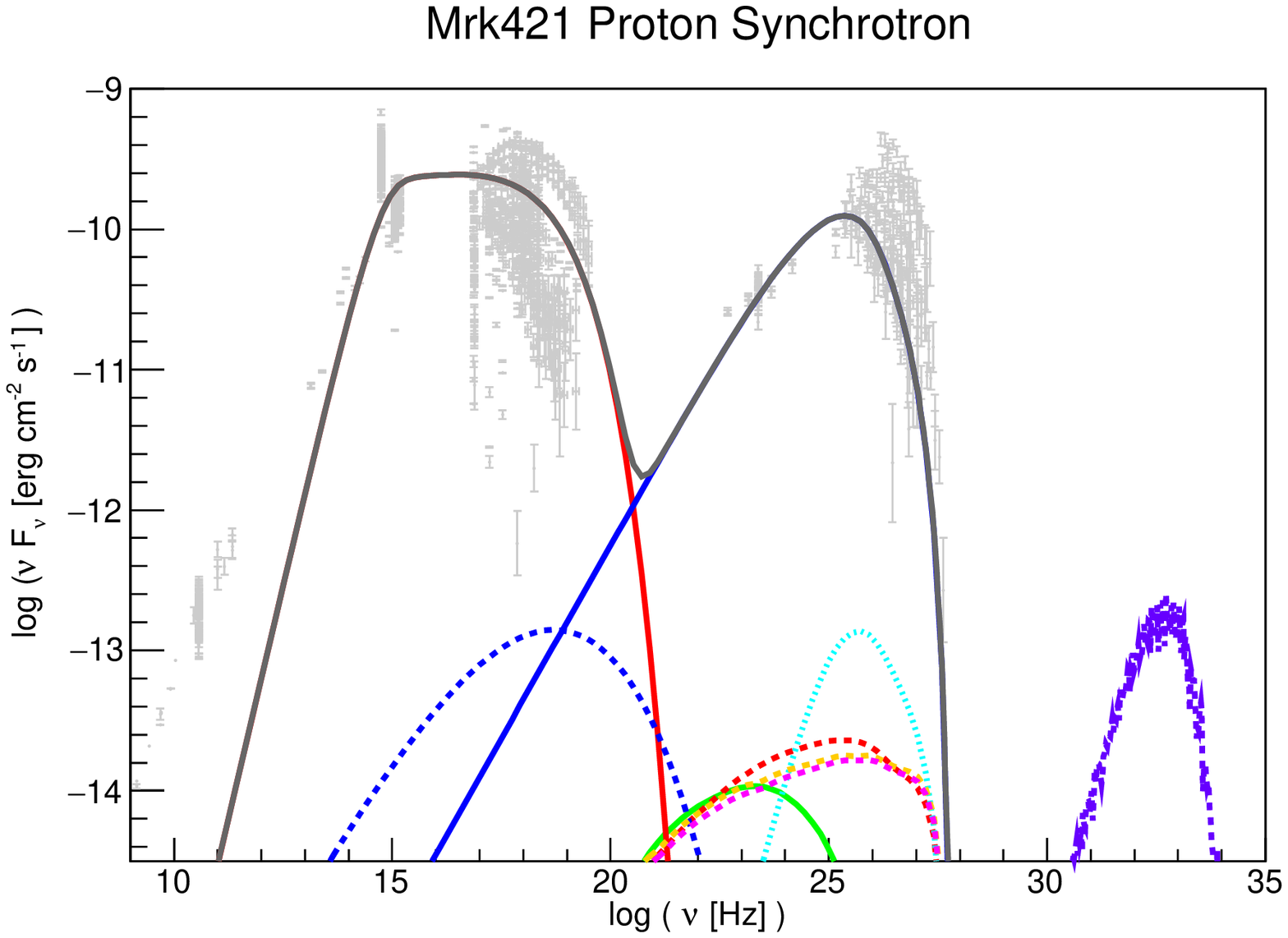}
\caption{Spectral energy distribution of the high-frequency-peaked BL Lacertae object Markarian 421, showing in gray archival data from the SSDC SED builder (\url{https://tools.ssdc.asi.it/SED/}). \textit{Top~left:} one-zone SSC modeling of the SED, with~in red synchrotron emission by electrons and in green synchrotron-self-Compton emission; model parameters are $\delta = 30$,  $R = 1.6\times10^{16}\ \textrm{cm}$, $B^\prime =40\ \textrm{mG}$, $\gamma^\prime_{e,min} = 100$, $\gamma^\prime_{e,break} = 6\times10^{4}$, $\gamma^\prime_{e,Max} = 10^{6}$, $n_{e,1} = 1.8$, $n_{e,2} = 3.5$, $L_{e} = 5.1\times 10^{43}\ \textrm{erg s}^{-1}$. \textit{Top~right:} one-zone lepto-hadronic modeling of the SED, with~in blue synchrotron emission by protons, in~dotted-blue synchrotron emission by secondary pairs from proton–synchrotron, in~dotted-red synchrotron emission by the Bethe–Heitler cascade, in~dotted yellow synchrotron  emission by the $\pi^0$ cascade, and~in dotted pink synchrotron emission by the $\pi^\pm$ cascade; neutrino emission is in violet; model parameters are same as the first plot, with~in addition a proton distribution with $\gamma^\prime_{p,min} = 1$, $\gamma^\prime_{p, Max} = 2\times10^{8}$, $n_p = 1.8$,  $L_{p} = 1.3\times 10^{47}\ \textrm{erg s}^{-1}$  \textit{Bottom:} one-zone hadronic modeling of the SED; model parameters are $\delta = 30$,  $R = 2.5\times10^{15}\ \textrm{cm}$, $B^\prime =50\ \textrm{G}$, $\gamma^\prime_{e,min} = 300$,  $\gamma^\prime_{e,Max} = 3\times10^{4}$, $n_{e} = 2.9$, $L_{e} = 5.4\times 10^{39}\ \textrm{erg s}^{-1}$, $\gamma^\prime_{p,min} = 1$, $\gamma^\prime_{p, Max} = 3\times10^{9}$, $n_p = 1.9$,  $L_{p} = 4.7\times 10^{43}\ \textrm{erg s}^{-1}$.  \label{figmrk421}}
\end{figure}

A one-zone SSC model can describe well the SEDs of HBLs (assuming that the radio emission comes from a larger, less dense region in the jet). An~example of an SSC modeling of the HBL Markarian 421 is shown in the top-left panel of Figure~\ref{figmrk421}. In~the recent years, a~number of algorithms have been published to fully constrain the SSC model parameter space, and~have been applied successfully to HBLs \citep{Finke08, Man11, Zhang12, SSCconstraints, Ahnen17}. Typical best-fit solutions are:
\begin{linenomath}\begin{equation}
\begin{array}{l}
    \delta \simeq 10\textrm{--}50 \\
    R \simeq 10^{16\textrm{--}17}~\textrm{cm}\\
    B^\prime \simeq 10\textrm{--}100~\textrm{mG} \\
    \gamma^\prime_{min} \leq 10^{2-3} \\
     \gamma^\prime_{break} \simeq 10^{3\textrm{--}4} \\
    \gamma^\prime_{Max} \geq 10^{5\textrm{--}6} \\
    n_1 \simeq 1.5\textrm{--}2.5 \\
    n_2 \simeq 3.5\textrm{--}4.5 \\
    L_{e} = 10^{43-44}\ \textrm{erg s}^{-1}  
  \end{array}
\end{equation}\end{linenomath}
\textls[-15]{{where $L_e = 2\pi R^2c\Gamma^2u^\prime_e$ {is the electron luminosity in the observer's frame, and}~$u^\prime_e~=~m_ec^2 \int d\gamma_e^\prime \gamma_e^\prime N(\gamma_e^\prime)$ }is the electron energy density.} The SSC scenario fails to reproduce the SED of other blazar sub-classes: it~cannot reproduce the SEDs of LBLs, nor FSRQs; it can reproduce the SEDs of EHBLs at the expenses of a high Doppler factor ($\delta \geq 50$) and of a high value of $\gamma^\prime_{min} \geq 10^3$ \citep{Biteau20, Cerruti15}.

\subsubsection{Synchrotron-Self-Compton: Time~Signatures}

\textls[-19]{Now that we have a model which is able to describe the blazar SEDs (at least for HBLs), we can add the time information to further test it and constrain the parameter values. As~for the synchrotron case, we~can compute the evolution of the SED with time by solving Equation~(\ref{equationdiffeq}) (including inverse-Compton losses) and calculating both synchrotron and SSC emission as a function of time. One difficulty of SSC time-dependent modeling is that the inverse-Compton cooling term depends on the density of the synchrotron photons, which depends on the integral of the electron distribution itself as a function of time (which is what we want to solve). In~presence of significant SSC losses the differential equation becomes thus non-linear and the solution is significantly more complex~\citep{Zacharias10, Zacharias14}}.  Blazar~light-curves in the $\gamma$-ray band have been successfully fit with the SSC model \citep{2155HESS, Chen11}. Similarly~to synchrotron radiation, time-lags and hysteresis are also expected in SSC light-curves due to the energy dependency of both the acceleration and cooling terms \citep{Perennes20}. But~contrarily to what is observed in X-rays, there has not been any significant detection of time-lags or of hysteresis in the $\gamma$-ray band \citep{Albert07, Abey17}. {This~negative result is likely due to the intrinsic lower sensitivity of gamma-ray telescopes compared to X-ray ones, and~not to a failure of the model.}

Another important prediction of the SSC scenario is the correlation between the variability of the two SED components, which are expected to vary in concert. If~the variability is simply due to an achromatic change in the particle density in the emitting region, the~SSC component is expected to vary as the square of the synchrotron component (due to both the change in the particle distribution, and~in the synchrotron photons). A~correlation plot between X-rays and $\gamma$-rays is thus expected to show a simple quadratic relation. If~on the other hand the variability is energy-dependent, the~expected correlations are more complex, and~depend on the energy bands of the observations, the~SED peak frequencies, the~energy-dependency of the injection \citep{Katarzynski05}. Correlation studies among different wavebands are indeed a powerful tool to test radiative models when the statistics in an individual band does not allow light-curve fitting \citep{Fossati08, Acciari11, Aleksic15, Furniss15, Balokovic16, Gonz19}.

\subsubsection{External~Inverse-Compton}

\begin{figure}[H]
\centering
\includegraphics[width=10cm]{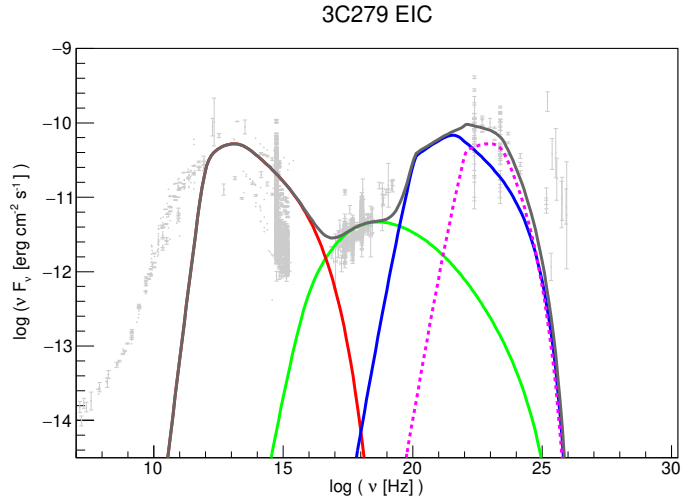}
\caption{Spectral energy distribution of the flat spectrum radio quasar 3C279, showing in gray archival data from the SSDC SED builder (\url{https://tools.ssdc.asi.it/SED/}). One-zone external-inverse-Compton modeling of the SED, with~in red synchrotron emission by electrons; in green synchrotron-self-Compton emission; in blue EIC emission over dust photons; and in pink EIC emission over Ly$\alpha$ photons. Parameters are $\delta = 30$,  $R = 1.3\times10^{16}\ \textrm{cm}$, $B^\prime =1.5\ \textrm{G}$, $\gamma^\prime_{min} = 50$, $\gamma^\prime_{break} = 350$, $\gamma^\prime_{Max} = 10^{4}$, $n_1 = 2.5$, $n_2 = 3.5$, $L_{e} = 7\times 10^{44}\ \textrm{erg s}^{-1}$, $u^\prime_{Ly\alpha} = 1\ \textrm{erg cm}^{-3}$, $u^\prime_{torus} = 0.7\ \textrm{erg cm}^{-3} $  \label{fig3c279}}
\end{figure}

Synchrotron photons are not the only low-energy photons that can be up-scattered to $\gamma$-rays. The~SMBH environment is very bright, and~there is a huge variety of photon fields that can go through inverse-Compton scattering with electrons in the jet. This radiative process is usually called \textit{External-inverse-Compton} (EIC) and depends strongly on the properties of the external field and on the exact location of the emitting region $r$ (measured from the SMBH). The~computation of the EIC component is similar to the SSC one, but~requires as first step the transformation of the external photon field distribution into the jet's frame. For~blazar modeling, the~relevant photon fields, starting from the AGN core outwards, are:

\begin{itemize}
[leftmargin=*,labelsep=4.9mm] 
\item the SMBH accretion disk is a reservoir of thermal photons. EIC emission over disk photons is however hindered due to the Doppler deboosting \citep{Dermer93}: the plasma in the jet is travelling away from it with Lorentz factor $\Gamma$, and~thus the disk photon field energy density, in~the reference frame of the plasmoid, is strongly suppressed by a factor $\Gamma^{-3}$. EIC scattering over disk photons can thus dominate the overall $\gamma$-ray emission only if the emitting region is located close to the disk itself. In~this case the plasma can see part of the disk at angles large enough to reduce the de-boosting.

\item the BLR produces bright emission lines (Ly$\alpha$ being the dominant one) that can serve as target photon field for EIC emission \citep{Sikora94}. In~this case the emission is boosted or deboosted in the reference frame of the emitting region as a function of $r$ and $r_{BLR}$. It is important to underline that the BLR is structured, and~that different emission lines are produced at different distances $r_{BLR, i}$ from the SMBH. The~size of the BLR can be expressed as a function of its luminosity (or~other proxies such as the luminosity of the continuum at $5100$ \r{A}, or~the disk luminosity \citep{Greene05, Kaspi00}). The~EIC emission is thus a superposition of several components coming from the different lines \citep{Finke16}, with~their relative strength depending on $r$. A~second radiative component coming from the BLR is the photon field from the accretion disk which can be Thomson-scattered by electrons in the BLR and thus seen boosted in the reference frame of the blob. This component depends also on the optical depth $\tau_{BLR}$ of the BLR. A~parameter which is not so well known is the aperture angle $\alpha_{BLR}$ of the BLR: it can have a spherical geometry ($\alpha_{BLR} = 0$) or be flattened over the disk ($\alpha_{BLR} \rightarrow \pi/2$)~\citep{Tavecchio12, Lei14}. The~assumed BLR geometry has a direct impact on the photon field seen in the blob reference frame, and~in the limit of a flat BLR the computation of the associated EIC component becomes similar to the one from the accretion~disk. 

\item the emission from the dusty torus is thermal and can be well described with a single-temperature black-body distribution with temperature $T \simeq 1000$ K . The~torus location can also be parametrized as a function of the disk luminosity $r_{torus} = 2.5\times10^{18} \sqrt{L_{disk} / 10^{45}~\textrm{erg s}^{-1}}$ cm \citep{Sikora09}. Similarly to the BLR aperture angle, the aperture angle of the torus is also not well known. It is commonly assumed to be of around $\pi/4$.  As with the BLR photon field, the~blob of plasma sees a constant photon density from the torus for $r \ll r_{torus}$,  and~then a rapidly decreasing photon field while crossing  $r_{torus}$ and leaving the torus behind \citep{Blaz00}. An~example of an EIC modeling of the FSRQ 3C279 is shown in~Figure~\ref{fig3c279}.

\item A ubiquitous photon field is the one from the Cosmic Microwave Background (CMB): electrons~in the jet can also scatter these photons to higher energies. This EIC emission has the important properties of being the only one dependent on the redshift of the source. EIC over the CMB is not so efficient when the emitting region is located close to the SMBH, because~there are always {stronger} photon fields. But~it can become an important radiative process to explain X-ray and $\gamma$-ray emission from SMBH jets at larger scales (\citep{Tavecchio00}, but~see \citep{Meyer17}). l{Since the CMB intensity scales as} $(1+z)^4$, this emission process can play a more important role for high-redshift quasars~\citep{Schwartz02}.

\item as we discussed before, structured jets can explain the synchrotron emission from radio up to optical-X-rays. The~emission from the other regions of the jet can thus serve as target photon field for EIC scattering. In~configurations such as the spine-sheath scenario (a fast inner jet surrounded by a slower layer), the~emission from the external jet can be boosted in the $\gamma$-ray emitting region due to the relative motion of the two components \citep{Ghisellinispine, Hervet15, MacDonald15}.

\end{itemize}

\vspace{-10pt}

\subsubsection{Energy Budget of the Emitting~Region}

An important characteristic of all radiative models is its energy budget: how much energy is required to reproduce the observations? and how is energy distributed among the various components?  The total jet power (in erg s$^{-1}$) can be expressed as an explicit function of the various energy densities~as
\begin{linenomath}\begin{equation}
L_{jet} = 2\pi R^2c\Gamma^2(u^\prime_B + u^\prime_e + u^\prime_{ph})
\end{equation}\end{linenomath} 
where the factor of 2 takes into account that there are always two~jets. 

Equipartition between the various radiative components is always attractive especially in physical systems at equilibrium, because~it directly provides a minimum power solution. It is thus interesting to note that SSC solutions for HBLs are \textit{{not}} near equipartition, the~emitting region being always particle dominated \citep{Kino02, Tavecchio16}. On~the other hand, EIC solutions for FSRQs show near-equipartition conditions between particles, magnetic field, and~external  photons \citep{equip1}.

\subsection{Electron-Positron Pair~Production}

The last leptonic process that is relevant to calculate the radiative output of SMBH jets is pair-production via photon–photon interaction \citep{Gould67}
\begin{linenomath}\begin{equation}
\gamma + \gamma \rightarrow e^+ + e^- 
\end{equation}\end{linenomath}

This process is the inverse of pair annihilation, in~which one electron and one positron annihilate into 0.511 MeV photons. Pair-production can occur only if the energy of the two photons is higher than the rest mass of the leptons that are produced, which is $\varepsilon_1 \varepsilon_2 = h\nu_1 h\nu_2/m_e^2c^4 > 1$. The~cross-section of the interaction is \citep{Coppi90}
\begin{linenomath}\begin{equation}
\sigma_{\gamma\textrm{-}\gamma} (s) = \frac{3\sigma_T}{16}(1-s^2)\left[2s(s^2-2) + (3-s^4)\ln\left( \frac{1+s}{1-s} \right)  \right]
\end{equation}\end{linenomath}
where $s=\varepsilon_1 \varepsilon_2 (1-\cos{\vartheta}) / 2$, and~$\vartheta$ the interaction angle. The~cross-section can be reasonably well approximated as
\begin{linenomath}\begin{equation}
\sigma_{\gamma\textrm{-}\gamma} (\varepsilon_1, \varepsilon_2) \simeq \frac{2}{3} \sigma_T \delta \left(\varepsilon_2 - \frac{2}{\varepsilon_1} \right) \frac{1}{\varepsilon_1}
\end{equation}\end{linenomath}

This approximation is known as the $\delta$-approximation. The~pair-production process is more efficient when the energy of the two photons are inversely proportional. The~two photons disappear in the interaction, and~while for low-energy photon fields the disappearance of a single photon is negligible, for~the high-energy photon field a single photon can contribute significantly to the overall energy flux. For~$\gamma$-rays, pair-production can result in significant absorption, which can be expressed via the following absorption coefficient (similarly to Equation~(\ref{selfabscoef}); {note that the following equation is correct only if an angle-averaged cross-section is used, otherwise an additional integration over the angle is required}):

\begin{linenomath}\begin{equation}
\mu^\prime_{\gamma\textrm{-}\gamma} (\nu^\prime) = \int_{\varepsilon^\prime_{min}} ^{\varepsilon^\prime_{Max}} d\varepsilon^\prime\ n_{ph}^\prime (\varepsilon^\prime) \sigma_{\gamma\textrm{-}\gamma} (\varepsilon^\prime, h\nu^\prime/m_ec^2)
\end{equation}\end{linenomath}
where $n_{ph}^\prime (\varepsilon^\prime)$ is the number density of the low-energy photon field, which can be computed from the~intensity 

\begin{linenomath}\begin{equation}
 n_{ph}^\prime (\varepsilon^\prime) = \frac{4\pi}{hc} \frac{I^\prime_{\varepsilon^\prime} (\varepsilon^\prime) }{\varepsilon^\prime}
\end{equation}\end{linenomath}

The most relevant $\gamma$-$\gamma$ absorption term for $\gamma$-ray blazars is external to the source, and~it is represented by the integrated emission from all galaxies (stars and dust) that permeates the Universe, the~Extragalactic Background Light (EBL). The~EBL acts as an absorber for TeV photons in the Universe, and~this absorption effect increases with the distance from Earth and with the energy of the $\gamma$-ray photon \citep{EBL}. The~result is a softening of the spectrum of TeV blazars, and~an horizon for TeV astronomy. The~TeV horizon is defined as the energy-dependent redshift for which the opacity $\tau_{EBL} = 1$. Although~the EBL significantly hinders TeV observations of distant blazars (the most distant source observed in the TeV band with Cherenkov telescopes is at z = 0.944, although~detected only up to 0.175~TeV~\citep{0218}), this~absorption effect can be used to measure the EBL itself, opening up the path to $\gamma$-ray cosmology~\citep{Biteau15}. The~pairs produced in the interaction with the EBL are not lost, and~can produce further emission along the line of sight mainly via inverse-Compton scattering over CMB photons, although~energy losses from plasma instabilities are also investigated \citep{Broderick12}. This emission can be detected both as a new radiative component in the GeV band \citep{Plaga95}, or~as extended gamma-ray pair halos~\citep{Aharonian94}, both modulated by the strength of the magnetic field seen by the pairs. None of these features has been detected as per today~\citep{HESSpairhalos, VERITASpairhalos, Fermipairhalos}, and~these observations can be used to put constraints on the inter-galactic magnetic field~\citep{Neronov10, Taylor11}.

Going back to the inverse-Compton scattering in SMBH jets, all photon fields which are inverse-Compton scattered to $\gamma$-ray energies also act as an absorber to the same $\gamma$-ray photons via pair-production. Let us start by looking at the SSC case. Under~the $\delta$-function approximation for $\sigma_{\gamma\textrm{-}\gamma}$, and~by forcing an equality to the causality relation (Equation (\ref{causalityequation})), it is possible to write $\tau_{\gamma\textrm{-}\gamma} < 1$, where the pair-production opacity depends only on the Doppler factor $\delta_{SSC}$ (and physical observables)~\citep{Dondi95}, and~it is thus possible to provide a lower limit on $\delta_{SSC}$
\begin{linenomath}\begin{equation}
\label{equationdondi}
\delta_{SSC} \geq \left[ \frac{\sigma_T}{5hc^2} d_L^2 (1+z)^{2\alpha}\ \frac{F_{\nu}(\nu_0)}{\tau_{var}}  \right]^{1/(4+2\alpha)}
\end{equation}\end{linenomath}
where $\nu_0 = \left( 1.6 \times 10^{40} / \nu_{\gamma} \right) $ Hz is the frequency of the low-energy photons that act as primary absorber for photons at $\nu_{\gamma}$, and~$\alpha$ is the index of the flux density $F_{\nu}$ at $\nu_0$. For~typical blazars parameters, Equation~(\ref{equationdondi}) results in  $\delta_{SSC} \geq 5-10$. The~detection of TeV photons from HBLs can thus be used to put a strong limit on  $\delta$, constraining further the parameter space of the SSC~model. 

\subsubsection{On the Location of the $\gamma$-ray Emitting Region in~FSRQs}

For the EIC case, the~opacity to $\gamma$-$\gamma$ pair production depends on the location $r$ of the emitting region with respect to the external fields. The~opacity is expressed as the integral from $r$ outwards of the absorption coefficient:
\begin{linenomath}\begin{equation}
\tau_{\gamma\textrm{-}\gamma} (\nu^\prime) = \int_r^\infty dr^\prime\ \mu^\prime_{\gamma\textrm{-}\gamma} (\nu^\prime)
\end{equation}\end{linenomath}

The presence of the BLR is very relevant for opacity calculations: if $r<r_{BLR}$, radiation from the BLR will significantly absorb $\gamma$-ray photons. Assuming that the most relevant emission line is Ly$\alpha$ (13.6~eV), this will cause an absorption feature at $\simeq$100 GeV. Detailed calculations including all relevant BLR lines and the thermal continuum from the disk scattered by the BLR \citep{GhiselliniMadau, Donea03, Liu06, Ghisellini09, Finke16}  show that the very detection of $\gamma$-ray photons with energies greater than 100 GeV from FSRQs is enough to put the location of the emitting region beyond the BLR. Observations of FSRQs at $E>100$ GeV confirmed this scenario and proved that, at~least for the sources that have been detected at high energies, the~emitting region is at or beyond $r_{BLR}$ \citep{MAGIC1222,Nalewajko14, 1441MAGIC, 1441VER, Costamante18, HESS3C279, HESS0736}.

The dusty torus also leaves an absorption imprint on the $\gamma$-ray spectrum. Given its typical temperature of $\simeq$1000 K, the~pair-production absorption on the torus photons is expected to be seen in the TeV band in $\gamma$-ray FSRQs. Due to the simultaneous absorption on the EBL, and~to the fact that FSRQs have an SED peak in the 100-MeV range with very few TeV photons, such a cut-off has never been observed as per today \citep{Galanti20}.

\subsubsection{Pair~Injection}

The electron-positron pairs that have such an important role for $\gamma$-ray absorption in SMBH jet are injected in the emitting region, and~can be treated as a secondary injection of leptons. They will behave exactly as the primary ones, and~will radiate synchrotron and inverse-Compton photons, reaching~their own equilibrium distribution. The~big advantage is that while we do not know the  details of the primary accelerator and we simply parametrized it, we have the information on the energy distribution of the pairs injected via pair-production \citep{Boettcher97}:

\begin{linenomath}\begin{equation}
\begin{array}{l}
Q^\prime(\gamma^\prime_e) = \frac{3\sigma_T}{32} \int_{\gamma^\prime_e}^\infty d\varepsilon_1 \frac{n_{ph}^\prime (\varepsilon_1)}{\varepsilon^3_1} \int_{\frac{\varepsilon_1}{4\gamma^\prime_e(\varepsilon_1 - \gamma^\prime_e)}}^\infty d\varepsilon_2 \frac{n_{ph}^\prime (\varepsilon_2)}{\varepsilon^2_2} \times \\
\times \left[\frac{4\varepsilon^2_1}{\gamma^\prime(\varepsilon_1 - \gamma^\prime)} \log\left( \frac{4\gamma^\prime_e \varepsilon_2 (\varepsilon_1 - \gamma^\prime)}{\varepsilon_1} \right) - 8\varepsilon_1 \varepsilon_2 + \frac{2\varepsilon^2_1(\varepsilon_1 \varepsilon_2 -1)}{\gamma^\prime(\varepsilon_1 - \gamma^\prime)}  - \left(1 - \frac{1}{\varepsilon_1 \varepsilon_2} \right) \left(\frac{\varepsilon^2_1}{\gamma^\prime(\varepsilon_1 - \gamma^\prime)} \right)^2  \right]
\end{array}
\end{equation}\end{linenomath}
where $n_{ph}^\prime (\varepsilon_1)$ and $n_{ph}^\prime (\varepsilon_2)$ are the photon densities of the two photon fields. The~injection term can then be used to compute the steady-state distribution and the associated synchrotron and inverse-Compton emission following Equation~(\ref{equationdiffeq}). For~single-zone blazar leptonic models, emission by this secondary leptonic population is usually negligible,  hidden by the primary~one.

\section{Hadronic Radiative~Processes}
\unskip
\vspace{-10pt}
\subsection{Proton–Synchrotron~Emission}
Let us now consider the case of a jet in which there are also protons (and possibly higher-Z nuclei) together with leptons. As with electrons, the~proton energy distribution is defined between $\gamma^\prime_{p, min}$ and $\gamma^\prime_{p, Max}$ as a power-law function:
\begin{linenomath}\begin{equation}
N^\prime_p (\gamma^\prime_p) = N^\prime_{0,p} \gamma^{\prime -n_p}_{p}
\end{equation}\end{linenomath}

 Going back to our spherical plasmoid in the jet, the~first radiative process that needs to be studied is synchrotron radiation by protons.With respect to electrons, the~emission from protons is suppressed and needs to be compensated by increasing either the magnetic field value, the~particle density, or~both. It is interesting to test the possibility that proton–synchrotron radiation is responsible for the high-energy SED component in blazars. Indeed, proton–synchrotron radiation would have the same index $\alpha_p = (n_p - 1)/2$ as SSC radiation, assuming that the injection index of protons is similar to the one of electrons. Following Equation~(\ref{synpeak}) (replacing the electron {mass} with the proton mass) it is possible to have a proton–synchrotron peak frequency at around $10$-$100$ GeV, consistent with HBL SEDs, assuming $\gamma_{p, Max}^\prime \simeq 10^9$, $B^\prime \simeq 10$--$100$ G, and~$\delta = 10$--$50$. The~proton–synchrotron solution for HBLs lies thus in a completely different part of the parameter space with respect to the SSC solution, with~a much stronger (a thousand times higher) magnetic field \citep{Mannheim93, Aharonian00, Mucke01}. The~number of free parameters of the proton–synchrotron scenario is much larger than the SSC one, due to the fact that the proton distribution adds 5 additional free parameters (assuming an additional $\gamma^\prime_{p, break}$ due to cooling, and~the two indexes $n_{p, 1/2}$), for~a total of 14 model parameters. Such a model is clearly degenerate, although~the number of free parameters can be considerably reduced by assuming some physically motivated constraints, such as: self-consistency of break Lorentz factors, co-acceleration of electrons and protons (and thus equality of $n_{e,1}$ and $n_{p,1}$), constraint on $\gamma_{p, Max}$ via equation of acceleration and cooling, constraint on $R$ via causality arguments. But~even imposing all these constraints the model remain degenerate, and~{and we can only constrain allowed regions in the parameter space} \citep{Cerruti15}.

\subsection{Proton-Photon~Interactions}

Protons in the jet interact with the low-energy photon fields, both internal (i.e., the electron synchrotron radiation) and external. But~contrarily to electrons, for~which the dominant interaction is inverse-Compton scattering, the~dominant processes for protons are photo-meson production and Bethe–Heitler pair-production. Inverse-Compton scattering for protons does happen, but~it can be safely neglected for the modeling of AGN~emission.

\subsubsection{Photo-Meson~Production}
The photo-meson (or photo-pion) process is the production of pions in proton-photon interactions:
\begin{linenomath}\begin{equation}
\begin{array}{l}
p + \gamma \rightarrow p^\prime + \pi^0 \\
p + \gamma \rightarrow n + \pi^+  \\
p + \gamma \rightarrow p^\prime + \pi^+ + \pi^-  \\
\end{array}
\end{equation}\end{linenomath}

The photo-meson cross-section is dominated by resonances ($\Delta$ and $N$ baryons), direct pion production and multiple pion production.  For~astrophysical applications, the~relevant information is the production of pions, both neutral and charged. In~order to produce pions, the~photon energy in the proton frame has to be higher than $m_\pi c^2 + (m_\pi/2 m_p) c^2 \simeq 145$ MeV. In~the emitting-region frame, $\epsilon^\prime \gtrsim (145/\gamma^\prime_p)$  MeV, meaning that protons with Lorentz factor $\gamma^\prime_{p} \geq 10^{7}$ can photo-meson-produce over ultra-violet~photons.

Pions then decay primarily ($\sim 100\%$ branching ratio) as
\begin{linenomath}\begin{equation}
\begin{array}{l}
\pi^0 \rightarrow 2\gamma \\
\pi^+ \rightarrow  \mu^+ + \nu_\mu \rightarrow e^+ + \nu_e + \bar{\nu}_\mu + \nu_\mu  \\
\pi^- \rightarrow  \mu^- + \bar{\nu}_\mu  \rightarrow e^- + \bar{\nu}_e +  \nu_\mu + \bar{\nu}_\mu \\
\end{array}
\end{equation}\end{linenomath}

Photo-meson production has thus one key property: neutrinos are produced together with photons, and~they can escape the emitting region without suffering from any additional absorption or energy losses. Detection of neutrinos from an AGN is thus a smoking-gun signal for the presence of relativistic protons in the jet. This implies that AGNs can accelerate protons to high energies, and~can be responsible for the flux of particles that we call, once they reach Earth, \textit{cosmic rays}. Joint neutrino and photon observations of AGNs can be used to constrain three key parameters of the model: how~many protons are accelerated, what is their energy distribution, and~what is their maximum~energy. 

Together with the neutrinos, photo-meson production injects photons and electron/positron pairs in the emitting region. The~spectrum of the injected pions has a maximum energy equal to the proton maximum energy but reduced by the inelasticity $\kappa \simeq 0.2$. The~maximum energy of photons from the $\pi^0$ decay can be as high as $0.1 \gamma^\prime_p$. For~protons with $\gamma^\prime_p = 10^7$, it translates into photons with $E \simeq 10$ PeV. These photons do not reach us: they are absorbed via pair-production both in the jet, and~in the path to Earth. The~absorption occurring in the source {produces} an electron/positron pair that starts radiating (and cooling). Given the high energy of the first photon, the~synchrotron radiation by these secondary leptons is at energies high enough that the synchrotron photons can suffer a second pair-production, and~so on. Photons from the $\pi^0$ decay can thus trigger a synchrotron-supported pair-cascade in the emitting region (or, if~the soft photon field energy density is larger than the magnetic one, an~inverse-Compton supported pair-cascade). Emission from the cascade can be computed calculating the pair distribution at equilibrium and their synchrotron radiation. When~$\tau_{\gamma\textrm{-}\gamma} \gg 1$ the cascade is \textit{saturated}, and~the resulting synchrotron emission is a power-law with index $\alpha = 1$, or~in terms of energy flux $p=0$: the energy transferred from protons to pions and then to high-energy photons, is then redistributed to low~energies.
 
Electrons and positrons produced in the decay of charged pions go through a process similar to the photons from $\pi^0$: they have extremely high-energy, and~their synchrotron radiation is also in the PeV band. They thus trigger as well a pair-cascade, which behaves exactly like {the} one from $\pi^0$. 

Another important emission process from photo-meson interactions is the synchrotron radiation by $\mu^\pm$. Although~not stable leptons, muons in the jet can survive long enough to radiate synchrotron photons before decaying to electrons/positrons. To~compute the muon energy distribution at equilibrium, the~muon decay term must be added to Equation~(\ref{equationdiffeq}) as follows:
\begin{linenomath}\begin{equation}
\label{equationdiffeqmu}
\frac{\partial}{\partial t} N^\prime_\mu(\gamma^\prime_\mu, t) = \frac{\partial}{\partial \gamma^\prime_\mu} \left[ \gamma^\prime_\mu \frac{N^\prime_\mu(\gamma^\prime_\mu, t)} {\tau_{syn}(\gamma^\prime_\mu)} \right] - \frac{N^\prime_\mu(\gamma^\prime_\mu, t)}{\tau_{ad}}  - \frac{N^\prime_\mu(\gamma^\prime_\mu, t)}{\gamma^\prime_\mu \tau_{dec}}  + Q^\prime_\mu(\gamma^\prime_\mu) 
\end{equation}\end{linenomath}
where $\tau_{dec} = 2.2 \times 10^{-6}$ seconds. The~corresponding integral solution for the muon distribution at equilibrium is:
\begin{linenomath}\begin{equation}
N^\prime_\mu(\gamma^\prime_\mu) = \exp\left[-\frac{\gamma^\prime_{\mu, break}}{ \gamma^\prime_\mu} - \frac{\gamma^\prime_{\mu, break}\tau_{ad} }{ 2\gamma^{\prime 2}_\mu \tau_{dec} } \right] \frac{\gamma^\prime_{\mu, break}\tau_{ad}}{\gamma^{\prime 2}_\mu} \int_{\gamma^\prime_\mu}^\infty d\zeta Q^\prime_\mu(\zeta) \exp \left[ \frac{\gamma^\prime_{\mu, break}}{\zeta} + \frac{\gamma^\prime_{\mu, break}\tau_{ad} }{ 2\zeta^2 \tau_{dec} } \right]
\end{equation}\end{linenomath}

Synchrotron radiation by muons emerges as a third radiative component, at~higher energies compared to the synchrotron radiation by the parent protons. In~some parts of the parameter space it can represent the dominant radiative emission in the TeV band. Another important consequence is that every time the muon synchrotron radiation is important, it means that muons can cool efficiently before decaying into electrons/positrons, and~this energy loss has to be included when calculating their injection in the emitting~region.

An interesting property of photo-meson interactions is the creation of neutrons, which can escape the emitting region without interacting with magnetic fields, nor producing Bethe–Heitler pairs (see Section \ref{bhpairproduction}). These~neutrons can transfer a significant amount of energy at much larger distances downstream the jet~\citep{Begelman90, Atoyan03}. In~presence of dense photon fields they can trigger additional photo-meson production, or~they can decay into protons (with a life-time of $\gamma_n \times 880$ seconds) and radiate again synchrotron photons in the presence of magnetic fields. This ``neutral beam'' model has drawn attention for its possibility to naturally produce two separate emitting regions in the jet, physically separated but casually connected~\citep{Dermer12, Zhang20}. 

The main problem we are facing with photo-meson production is that the exact computation of the energy distribution of all secondary particles injected in the emitting region is a complex numerical task. The~best approach is to compute them via Monte-Carlo simulations \citep{SOPHIA}. Several approaches to fit the results of Monte-Carlo simulations and provide a parametrization of the injected secondary particles over a wide parameter space are particularly useful \citep{Kelner08, Hummer}. 

\subsubsection{Bethe–Heitler~Pair-Production}
\label{bhpairproduction}
Bethe–Heitler pair-production is the following process
\begin{linenomath}\begin{equation}
p + \gamma \rightarrow p^\prime +  e^+ + e^- 
\end{equation}\end{linenomath}

It is a process which is in competition with the photo-meson production, although~it happens at lower energies. The~threshold for this interaction is lower than the photo-meson one by a factor $m_e / m_\pi \simeq 0.004$. In~terms of proton Lorentz factors, protons with $\gamma^\prime_{p} \geq 10^{5}$ can pair-produce over $10$~eV~photons.

For low-energy protons, the~Bethe–Heitler pair production is the main proton-photon interaction, but~as soon as the proton energy becomes greater than the energy threshold for photo-meson production, the~latter takes over. As~expected, pairs injected via this process have a lower energy compared to pairs from the photo-meson \citep{PetroMasti}. The~energy distribution of the injected Bethe–Heitler pairs can be computed analytically \citep{Blumenthal70, Chodorowski92, Kelner08}, or~using a Monte-Carlo approach \citep{Protheroe96, Masti05}. Their fate is the same as all other leptons in the emitting region: they radiate synchrotron and inverse-Compton photons, and~if their energy is high enough (and if the low-energy photon field is dense enough) they can trigger a~pair-cascade. 

\subsection{Hadronic and Lepto-Hadronic~Models}

Blazar hadronic models are a large family of models which share the presence of a parent population of protons that are accelerated in the SMBH jet. But~as discussed above, from~the electromagnetic point-of-view the only true hadronic radiative process is proton–synchrotron emission. Solutions in which the high-energy emission is produced by secondary leptons produced in p-$\gamma$ interactions are usually referred to as \textit{lepto-hadronic} models.

Pure hadronic (proton–synchrotron) models can fit blazar SEDs \citep{Mannheim93, Aharonian00, Mucke01, Bottcher13}, but~they face a major problem: the proton density required to fit the data is rather high, and~when expressed in terms of $L_p$ it gets close or clearly much higher than the Eddington luminosity of the SMBH \citep{Sikora09}. Although~$L_{Edd}$ should be treated as order-of-magnitude estimate for the available accretion power, if~the power in the jet required to fit the data is super-Eddington by several orders of magnitudes it means that either the relation between accretion and jet needs to be revisited, or, more likely, that this particular solution is not viable. This \textit{energy-crisis} for hadronic models is particularly true for FSRQs~\citep{Zdziarski}. For~low-luminosity HBLs and EHBLs it is not the case, and~hadronic solutions with $L_{jet} < L_{Edd}$ can be found \citep{Cerruti15, Petropoulou15}. Even if the high-energy SED component is dominated by proton–synchrotron radiation, emission by secondary leptons from p-$\gamma$ interactions can emerge (namely for compact and dense solutions) in the X-rays and in the TeV band \citep{Mucke03, Abdo11, Zech17}. In~terms of energy budget of the emitting region, proton–synchrotron solutions are out of equipartition, with~$u^\prime_B \geq u^\prime_p$. An~example of a hadronic modeling of the HBL Mrk 421 is shown in the second plot of Figure~\ref{figmrk421}.

In the lepto-hadronic solutions, the~proton–synchrotron radiative component is suppressed, and this can be easily achieved by reducing the magnetic field to $B^\prime \leq 1$ G. We are back in the part of the parameter space where the SSC emission dominates  and indeed the simplest lepto-hadronic model is an SSC solution loaded with relativistic protons that produce secondary leptons via p-$\gamma$ interactions over the electron synchrotron photon field (this solution is called \textit{one-zone} lepto-hadronic model, to~highlight the fact that all radiative mechanisms are coming from a single emitting region in the jet, and~there external photon fields are negligible). The~emission by secondary leptons emerges again in the X-rays (as Bethe–Heitler component) and in the TeV band (as photo-meson component) \citep{Petropoulou15}. This kind of lepto-hadronic solutions are typically more promising in terms of $\nu$ output than pure hadronic solutions. As with hadronic solutions, lepto-hadronic solutions also face energetic issues: especially if the goal of the modeling is to maximize the neutrino output, the~required jet power can quickly become very high. A~possible way-out to this issue is to use as target photon field not the electron synchrotron radiation, but~external photons. By~increasing the density of the target photon field, it is possible to achieve the same p-$\gamma$ output lowering the required proton power. The~discovery of the blazar TXS0506~+~056  as first neutrino blazar candidate \citep{TXS0506}, indicates that a lepto-hadronic solution over an external field may be the favored scenario \citep{Ansoldi18, Keivani18, Murase18, Gao19, Cerruti19, Righi19}. In~terms of energy budget of the emitting region, mixed lepto-hadronic are usually particle dominated, although~the {result} depends on the exact shape of the proton distribution (a soft/hard distribution has a significant impact on $u^\prime_p$, but~a marginal effect on the photon and neutrino emission). An~example of a mixed lepto-hadronic modeling of the HBL Mrk 421 is shown in the third plot of Figure~\ref{figmrk421}.

Hadronic radiative models are of particular interest any time leptonic ones face difficulties. One~of the most interesting cases is the unusual SED of the nearby radio-galaxy Centaurus A, which shows a unique third radiative component emerging at $\simeq$100 GeV \citep{CenAspectrum}. This additional component can be naturally explained in a hadronic or lepto-hadronic scenario \citep{PetropoulouCenA, Fraija, SahuCenA}.

Among blazar hadronic radiative models, it is worth mentioning another potential contribution produced not in the jet, but~rather in the path from the source to the observer. If~AGNs accelerate cosmic rays, a~hadronic beam can be launched emitting photons and neutrinos while travelling towards Earth \citep{Essey10, Murase12}. So far no evidence for this hadronic contribution has been detected. For~a recent study of propagation effects comparing both hadronic cascades and electromagnetic cascades (the~ones triggered by the absorption over the EBL) see \citep{Dzhatdoev17}.

\subsubsection{Hadronic Models: Time~Signatures}

From the spectral point of view both leptonic and hadronic radiative models provide similarly good fit to current electromagnetic observations, but~they can in principle be distinguished from their temporal behavior. 
Time-dependent hadronic models require the solution of a system of coupled differential equations of the kind:
\begin{equation}
\frac{\partial N^\prime_X (t,E)}{\partial t} = Q^\prime_X (t,E) - L^\prime_X (t,E)
\end{equation}
for each species $X$ (protons, photons, neutrinos, leptons), where $Q^\prime_X (t,E)$ and $L^\prime_X (t,E)$ represents the injection and loss terms, respectively. From~a numerical point of view, hadronic time-dependent codes are significantly more complex than leptonic ones, and~have been extensively developed only in the recent years \citep{Masti05, Dimitra12, Weidinger15, Diltz15}. As with leptonic models they can be tested on data in two main ways: fitting light-curve profiles, and~studying multi-wavelength~correlations.

For proton–synchrotron models, although~the two SED components are produced by two distinct particle populations, it is still possible to reproduce the observed correlations assuming that electrons and protons share the same acceleration mechanism \citep{Masti13}. On~the other hand, absence of correlations can also be reproduced and indeed, one of the best applications for blazar hadronic models, are~the so-called \textit{orphan flare}, bright $\gamma$-ray flares without multi-wavelength counterparts, which cannot be reproduced in a leptonic one-zone scenario \citep{Kraw04, Bottcher05, Sahu13}.

Time-dependent models have also the advantage to allow the study of non-linear effects in the development of the hadronic pair-cascades, in~which photons from the cascade become targets for pair-production and can result in a unique pulsating signature \citep{Petropoulou12}.

\subsection{Proton–Proton~Interactions}

The final hadronic process that needs to be taken into account is pion production in proton–proton interactions. As with the photo-meson production, a~highly relativistic proton interacts with a target low-energy proton. For~astrophysical applications, the~only relevant process is the production of neutral pions $\pi^0$ and $\eta$ mesons, which then decay into $\pi^\pm$. The~result of the interaction is~then 

\begin{linenomath}\begin{equation}
\begin{array}{l}
p + p \rightarrow \pi^0 + Y \\
p + p \rightarrow \pi^+ + \pi^- + Z\\
\end{array}
\end{equation}\end{linenomath}
where $Y$ and $Z$ are any other particle produced in the interaction.   The~decay of charged pions is responsible for the associated neutrino emission.  As with the photo-meson process, the~outcome of proton–proton interactions is better studied via Monte-Carlo simulations, although~useful parametrizations have been developed \citep{Kelner06, Kafe14}. 
In blazar's jets, proton–proton interactions represent usually a subdominant contribution. In~order to have a significant $p$-$p$ output a very proton-loaded jet is needed \citep{Reynoso11}. An~alternative scenario is represented by interactions of the jet with obstacles, such as BLR clouds, or~stars \citep{Barkov12, Barkov12b, BoschRamon12}.

\section{Polarization~Signatures}

An important observable that has the potential to uniquely determine the radiation mechanism, as~well as the physical properties of the jet, is polarization. A~specific signature of synchrotron radiation is indeed that the emission is linearly polarized. 
For a power-law electron energy distribution with index $n$, the~maximum polarization fraction is
\begin{linenomath}\begin{equation}
\Pi (n) = \frac{n + 1}{n +\frac{7}{3}}
\end{equation}\end{linenomath}
which is equal to $0.69$ for $n=2$. Detection of polarized radiation in blazars is thus a key prove that what is observed from radio to optical is synchrotron radiation \citep{Robopol}. The~polarization angle can be used to map the {orientation}  of the magnetic field in the jet. Recent multi-wavelength campaigns have shown not only that optical polarization is a characteristic of blazars, but~that both the polarization fraction and the polarization angle show drastic variability, often correlated with flares at other wavelengths \citep{Robopol2}.
For an updated review of optical polarization in AGN jets, see \citep{ZhangReview} in this special issue. Variability in the optical polarization can also be used to put constraints on the geometry of the emitting regions, and~in particular on the relative role of the small plasmoid (the single-zone that emits from optical to gamma-rays) and the larger jet \citep{Ulisses14,Magictwozones}. 

Inverse-Compton scattering is also expected to be polarized, depending on the polarization of the target photon field \citep{Krawczynski12}. X-ray and gamma-ray polarimetry have the potential to significantly constrain blazar radiative models: proton–synchrotron radiation is expected to be significantly much more polarized than SSC radiation or EIC radiation (which is no polarized at all if the external field is also not polarized). Leptonic and hadronic radiative processes can thus be distinguished using this unique observable \citep{ZhangBottcher, Paliya18}.

\section{Summary and~Perspectives}

Relativistic jets from supermassive black holes emit photons over the whole electromagnetic spectrum, from~radio to gamma-rays, and~since 2017 we have the first evidence that they might be bright in neutrinos as well. The~last decades have seen a significant improvement in our knowledge of their gamma-ray emission, and~the high-energy peak of their SED is now constrained as well as the low-energy one. These advances on the observational side have significantly constrained our models for the photon emission. The~state-of-the-art view is that the most likely emission process in FSRQs is external-inverse-Compton scattering over the bright photon fields around the SMBH environment. If~hadronic processes play a role, they should be subdominant in FSRQs, due to the large amount of power they require. For~HBLs the main radiation mechanism is SSC, or~EIC in a structured jet. In~this case, pure hadronic models (proton–synchrotron dominated) cannot be excluded looking at the jet energy budget. Observations of the first blazar neutrino candidate, TXS~0506~+~056, favor a mixed scenario with a leptonic dominated SED with subdominant hadronic components in the form of pair-cascades emerging in the hard-X-rays and the TeV band. X-ray and TeV observations helped identifying a population of blazars with extreme peak frequencies (in the hard-X-rays and the TeV band, respectively), now known as EHBLs: in this case as well it is not possible to disentangle the different radiative mechanisms, and~both leptonic and hadronic solutions provide good description of the data.  In~the upcoming future new observations will provide key constraints on radiative models, and~will significantly increase our understanding of the physics of relativistic jets. Neutrino astronomy is just in its {early} years and additional data from IceCube and its upgrades \citep{IceCubeGen2}, ANTARES~\citep{ANTARES}, and~KM3NET \citep{KM3NET} will provide new constraints on hadronic emission processes in AGN jets, whether~new neutrino AGNs will be detected, or~not. X-ray polarimetry will also impact significantly the field, providing unique constraints on the radiative mechanism responsible for the X-ray emission in AGNs. The~Imaging X-ray Polarimetry Explorer (IXPE) satellite \citep{IXPE} is expected to be launched in late 2021. Since they are the most common sources in the gamma-ray sky, answers to {the current open problems} and new questions on SMBH jets physics will come from the next generation gamma-ray observatories, CTA in the TeV band \citep{CTA, ScienceCTA}, and~future proposed MeV observatories, such as AMEGO \citep{AMEGO}.

\vspace{6pt} 



\funding{M. Cerruti has received financial support through the Postdoctoral Junior Leader Fellowship Programme from la Caixa Banking Foundation, grant n. LCF/BQ/LI18/11630012. Funding for this work was partially provided by the Spanish MINECO under project MDM-2014-0369 of ICCUB (Unidad de Excelencia 'Mar\'{i}a de Maeztu'). Silvia Cerruti is acknowledged for the AGN scheme in {Figure}
~\ref{figone}.}

\conflictsofinterest{The authors declare no conflict of interest. The~funders had no role in the design of the study; in the collection, analyses, or~interpretation of data; in the writing of the manuscript, or~in the decision to publish the results.} 

\abbreviations{The following abbreviations are used in this manuscript:\\

\noindent 
\begin{tabular}{@{}ll}
AGN & Active Galactic Nucleus\\
BLR & Broad Line Region \\
EBL & Extragalactic Background Light\\
EHBL & Extremely High-Frequency Peaked  BL Lac objects\\
EIC & External-Inverse-Compton \\
FR & Fanaroff–Riley (I and II)\\
FSRQ & Flat Spectrum Radio Quasar\\
HBL & High-Frequency Peaked  BL Lac objects\\
IBL & Intermediate-Frequency Peaked  BL Lac objects\\
$\Lambda$CDM & $\Lambda$-Cold-Dark-Matter\\
LBL & Low-Frequency Peaked  BL Lac objects\\
NLR & Narrow Line Region \\
SED & Spectral Energy Distribution \\
SMBH & Supermassive Black Hole\\
SSC & Synchrotron-Self-Compton\\
\end{tabular}}

\reftitle{{References}
}





\end{document}